\documentstyle[12pt,aaspp]{article}
\newcommand{\oxyfive}{O\(^{+5}\)}	% 5 times ionized oxygen
\newcommand{\nitfour}{N\(^{+4}\)}	% 4 times ionized nitrogen
\newcommand{\carthree}{C\(^{+3}\)}	% 3 times ionized carbon
\begin{document}

%\begin{abstract}
%\end{abstract}

\begin{center}
{\Large{\bf{Simulations of Supernova Remnants in Diffuse Media.
II. Three Remnants and Their X-Ray Emission}}}
\end{center}

\centerline{\large{{R. L. Shelton}}}

\centerline{\normalsize{{Laboratory for High Energy Astrophysics}}}
\centerline{\normalsize{{NASA Goddard Space Flight Center}}}
\centerline{\normalsize{{Greenbelt, MD  20771}}}
\centerline{\normalsize{{shelton@spots.gsfc.nasa.gov}}}
\centerline{\normalsize{{and}}}
\centerline{\normalsize{{Department of Physics and Astronomy}}}
\centerline{\normalsize{{Johns Hopkins University}}}
\centerline{\normalsize{{Baltimore, MD 21218}}}
\centerline{\normalsize{{shelton@pha.jhu.edu}}}

\vspace{1in}

\centerline{\Large{\bf{Abstract}}}

This paper provides detailed descriptions of the X-ray 
emission from supernova remnants evolving
in warm, low density, nonthermal pressure dominated regions
($T_o = 10^4$~K, $n = 0.001$~cm$^{-3}$, $P_{nt} = 1800$ or 7200
K~cm$^{-3}$).  Non-equilibrium ionization hydrocode
simulations are used to predict the high resolution spectra, 
$1/4$ and $3/4$~keV ROSAT PSPC countrates,
spatial appearance, color temperatures, and ratios of
O VII to O VIII emission line fluxes
as a function of time.

If undisturbed, the remnants are quite long lived,
surviving for 
$\sim 1.2$ to $\sim 1.6 \times 10^7$~yrs.  During their brief
energy conserving phases, the hot, highly pressurized gas 
behind their shock fronts copiously emit X-rays.  Thus, their
$1/4$~keV surface brightnesses are thousands 
%{\bf{(confirmed)}} 
of ROSAT PSPC 
counts~s$^{-1}$~arcmin$^{-2}$ and the remnants appear strongly
edge brightened.  
The onset of the radiative phase heralds the end of the extreme
X-ray luminosities,
but not the end of the X-ray emission.  
After the cool shell forms behind the shock front, 
the hot SNR bubble slowly radiates away the remaining 
energy, with a diminutive fraction released in the form of X-rays.
Thus, the
$1/4$~keV surface brightnesses are tens to hundreds of ROSAT PSPC 
counts~s$^{-1}$~arcmin$^{-2}$ and the remnants appear 
``centrally filled''.
The hot plasma within the remnant bubbles is always out of collisional
ionizational equilibrium.  Early on, the ionization states are much lower
than expected for such hot gas.  Later on, the ionization states are
higher.  

This paper also 
applies the standard observational analyses
for determining the color temperature, electron density, and thermal pressure
to ROSAT ``observations'' of one of the simulated remnants, thus providing a
map between observational results and physical conditions.
%The spectral color temperatures, calculated with the 
%ROSAT PSPC R1, R2, and R4 band response functions, 
%are $T_c \sim 10^6$ for the young supernova remnants and less
%for the older remnants.
%
% Paper III
%
% These simulations have been combined with the statistical distribution
% of isolated supernova progenitors in the Milky Way order to derive rough
% estimates.  If the high latitude supernova remnants have explosion
% energies of $E_o = 1.0 \times 10^{51}$ ergs and are evolving in a
% $P_{nt} = 7200$ K~cm$^{-3}$, $n_o \geq 0.01$ cm$^{-3}$ ambient medium, 
% then all of the southern halo's surface brightness can be attributed
% to isolated halo SNRs.  If the explosion energy or
% nonthermal pressure are halved, then the
% remnants produce fewer soft energy photons.
% In each case, bright young
% remnants should cover roughly one hundredth of the sky while older, 
% dimmer remnants should cover about half of the sky.  This
% is consistent with the patchiness seen at high southern latitudes.
% The simulations yield the fraction of
% volume filled by the warm or hot bubbles of supernova remnants, although
% the result depends
% on the height above the plane.  Bubbles between 290 and 660 pc, 
% fill up to $\sim10\%$ of the volume, while bubbles at greater heights
% fill smaller fractions of the volume.
The paper reports the
\oxyfive, \nitfour, and \carthree\ column densities 
for the simulated remnants.
The simulations may of interest to and are applied to studies of the 
Galactic halo and Local Bubble.  They may
also be of interest to studies of external galaxies and 
interarm regions of the Milky Way.

\noindent{\it{Subject Headings:}} supernova remnants -- 
ISM: diffuse X-rays -- 
ISM: \oxyfive --
ISM: \nitfour --
ISM: \carthree --
ISM: General --
Galactic: halo --
Galactic: Local Bubble --
Galaxies: external

\pagebreak

\section{Introduction} 

Researchers have been analyzing the X-ray emission
from supernova remnants for decades.
%The breadth of study is immense, ranging from
%detailed analyses of portions of individual SNRs to unresolved
%observations of possible SNRs in external galaxies.  
Often the results were compared with
analytical or computational models (for example, 
Hamilton and Sarazin 1984, Hughes and Helfand 1985, or the
Sedov-Taylor solution combined with ionizational 
collisional equilibrium spectral codes)
in order to determine the SNR's attributes or study SNR evolution.
%Many advances have been made via this technique.
%We have also been
%studying SNRs via analytical and computational methods for 
%decades, with results ranging from the very detailed simulations of
%portions of SNRs to the affects of populations of SNRs on a galaxy.
Much progress has been made with this approach.
Although the preponderant share of attention has been devoted to 
young remnants and those in moderate density environments, some
questions require models of older remnants and
remnants in very diffuse media.  For example,
how many soft X-ray photons should be attributed to 
a single SNR during its entire lifetime;
%so how bright is a population of
%SNRs in the halo of our Galaxy or in external galaxies?  
can SNRs
produce the observed surface brightness 
and spectrum of the Galactic halo or external
galaxies or must we consider other physical phenomena?
% SNRs in the
%halo of our Galaxy produce over their
%entire lifetimes?  From the answer, we can determine if 
%SNRs could be the sole sources of the observed halo surface
%brightness?  
What is the X-ray signature of a very old SNR and how does
it compare with the observed spectrum of the Local Hot Bubble or of other
quiescent X-ray regions?  What do the results tell us?

% those in the free expansion or energy conserving phases.  
Such questions cannot be adequately addressed with models of young
remnants alone because the more plentiful older remnants differ 
hydrodynamically and ionizationally.
In the young remnants, those in the free expansion and energy conserving 
phases, an energetic shock front sweeps up and heats up the ambient gas.
The temperature rises so quickly that
the degree of ionization does not equilibrate
with the gas temperature.  Thus the plasma is ``underionized'' and in
the process of ionizing.
Eventually, the shock weakens substantially and the recently swept up gas
is heated to much lesser temperatures.  In contrast to the
hotter gas in the center of the remnant, this 
gas quickly radiates its excess energy; the SNR is now
in the radiative phase.  
The SNR quickly evolves to have two distinct components, 
a ``shell'' of warm/cool gas
nearer to the shockfront and a bubble of hot gas nearer to the center of the 
remnant.  
%The hot bubble cools very slowly, at first via expansion and
%radiation and later by radiation alone.  
The hot bubble was once extremely hot and 
highly ionized.  During the remnant's expansion, this gas 
cooled faster than the atoms could recombine.
Thus the atoms have become ``overionized'' and are
in the process of recombining.

These questions require detailed models that span the remnants' lifetimes.
This paper was written to provide models for such inquiries.
This paper reports on the X-ray appearance of thermally 
conductive supernova remnants evolving in 
cool, non-thermal pressure dominated,
low density media.  The remnants were simulated with a
detailed hydrodynamic computer code coupled to
a nonequilibrium ionization code.  The paper provides 
detailed discussions of the remnants' physical states and atomic physics,
high resolution 50 to 1150~eV spectra, spatial maps, estimates of the
ROSAT PSPC $1/4$ and $3/4$~keV countrates, and 
\carthree, \nitfour, and \oxyfive\ column densities.

The paper is organized as follows:  Section 2 presents the
simulation techniques and model parameters for three remnants.
The primary hydrocode run (Remnant A) has 
an explosion kinetic energy ($E_o$) of $0.5 \times 10^{51}$~ergs,
ambient density ($n_o$) of $10^{-2}$~atoms~cm$^{-3}$, 
ambient temperature ($T_o$) of
$10^4$~K, and ambient non-thermal pressure ($P_{nt}$) of 1800~K~cm$^{-3}$.
Remnant B has four times the ambient non-thermal pressure and
Remnant C has four times the ambient non-thermal pressure and twice the
explosion energy.
%$E_o = 0.5 \times 10^{51}$~ergs, 
%$n_o = 10^{-2}$~atoms~cm$^{-3}$, $T_o = 10^4$ K, 
%and $P_{nt} = 7200$~K~cm$^{-3}$ and Remnant C has
%$E_o = 1.0 \times 10^{51}$~ergs, 
%$n_o = 10^{-2}$~atoms~cm$^{-3}$, $T_o = 10^4$ K, 
%and $P_{nt} = 7200$~K~cm$^{-3}$.
%The model parameters and
%simulation technique are presented in Section 2.
Section 3 treats Remnant A as a case study, thoroughly discussing its
%an example of a SNR in
%a tenuous, nonthermal pressure dominated environment,
%and discusses it in great detail.  Remnant A's
physical structure, evolution,  
%{\bf{(currently there is an ``overview to X-ray emission in there, too)}},
ionization states, spectra, total soft X-ray luminosity, spatial appearance
in soft X-ray emission, and \oxyfive, \nitfour, and \carthree\
content.
Sections 4 and 5 more succinctly discuss Remnants B and C.
Section 6 shows 
how the modeled emission would be interpreted if it were observed and
examined using popular observing instruments and analysis techniques.
These analyses include the determination of the color temperature,
electron density, thermal pressure, and O VII - O VIII diagnostics.  
Provisos are discussed in Section 7.
Section 8 applies the simulations to 
the Galactic halo and the Local Bubble.  Section
9 summarizes the major results.

Moreover, this paper, Paper I, and Paper III compose 
a larger study.  Paper I 
provides a very detailed examination of the high stage ions
(\oxyfive, \nitfour, and \carthree) in Remnant A - like SNRs.
Paper III combines the results of Papers I and II with the
predicted number of high latitude supernova progenitors then
compares the X-ray and high stage ion quantities and signatures 
with the observations.

\section{Modeling Method}

The supernova remnants are simulated using a Lagrangian mesh style
hydrocode, which employs non-equilibrium ionization and recombination,
thermal conduction, and non-thermal pressure.  The hydrocode
assumes spherical
symmetry and that the electrons and ions have equal kinetic temperatures.  
Paper I described the hydrocode in detail.  
The Raymond and Smith
(1977, 1993) code was used to calculate the non-equilibrium spectra
for the given ionizational and hydrodynamic states.  Between Paper I
and this paper, the implementation of the Raymond and Smith code was
improved.  The gas phase abundances of the ambient medium are taken as 
Grevesse \& Anders (1989) solar abundances.   
The simulations do not include additional metals from the progenitor.
Section 7 and Paper I discuss provisos on the model.

	Remnant A has an   
explosion energy ($E_o$) of $0.5 \times 10^{51}$~ergs, while 
the ambient medium has a temperature ($T_o$) of
$10^4$~K, ambient density of atoms ($n_o$) of $10^{-2}$~cm$^{-3}$, 
and ambient nonthermal pressure ($P_{nt}$) of $1.8 \times 10^3$~Kcm$^{-3}$.
Paper I also explains the
choice of these conditions.  Remnants B and C were made
with less conservative choices of the explosion energy and ambient
nonthermal pressure.  
Remnant B has $E_o = 0.5 \times 10^{51}$~ergs,
$T_o = 10^4$~K, $n_o = 10^{-2}$~cm$^{-3}$, 
and $P_{nt} = 7.2 \times 10^3$~Kcm$^{-3}$ and Remnant C has
$E_o = 1.0 \times 10^{51}$~ergs,
$T_o = 10^4$~K, $n_o = 10^{-2}$~cm$^{-3}$, 
and $P_{nt} = 7.2 \times 10^3$~Kcm$^{-3}$.

\section{Remnant A:
$n_o = 0.01$~cm$^{-3}$, $P_{nt} = 1.8 \times 10^3$, and
$E_o = 0.5 \times 10^{51}$~ergs}

\subsection{Remnant A -- Physical Structure}

The temperature, density, 
and pressure of the simulated SNR at $10^4$~yrs of age are shown as solid
lines in Figures~\ref{temperature}a, \ref{density}a, and
\ref{pressure}a, while the later times are shown in other linestyles.
The explosion creates an extremely hot, rarefied cavity enveloped by a 
hot, dense, rapidly expanding rim of swept up 
material.  
Thermal conduction has reduces the gradients in
temperature and density.  
%Without this effect, the central temperature
%(density) would be orders of magnitude greater (lower) but the
%pressure would be similar.
%Preliminary
%explorations using a denser ambient medium showed that the interior
%can be made quasi-isothermal even if thermal conduction 
%(or another process which mixes the thermal energy across large scales) 
%is allowed for only the first $\sim\frac{1}{5}$ of the time required
%for the SNR to form a cool shell.  Because the SNR evolves in a
%self similar fashion until radiative cooling becomes important, once
%established, the thermal structure endures during the adiabatic phase.

The blast wave or energy conserving phase of SNR evolution begins
when the expanding shock front has swept up much more mass than was
ejected in the supernova explosion.  This takes place within roughly 
$5 \times 10^4$~yrs.  During this phase,
the remnant continues to expand rapidly.  The thermal energy of the 
hot bubble is conserved and so as the remnant expands, 
the interior temperature declines.  Between 
$10^5$ and $2.5 \times 10^5$~yrs, the dense rim begins to 
radiate copiously and thus cool quickly.  By $5 \times 10^5$~yrs,
it has cooled to the ambient temperature, creating a spatially thick, cool
shell (see Figures~\ref{temperature}a and \ref{density}a).  
The shell is still bounded by a shock front which sweeps up
and thermalizes the gas it encounters.  Now, however, the shock is weaker
and the post shock temperature is less than $10^5$~K.  
%This gas is not X-rays emissive.
The SNR is now in its snow plow or momentum conserving phase.

In time, the shell widens because the shock front expands much faster than
the hot bubble.  The hot bubble (also called the interior) 
continually erodes as the hot gas
nearest its periphery cools and adds to the cool shell.  In addition,
the ambient pressure restrains, and, at about $2 \times 10^6$ yrs,
halts the bubble's expansion.  Subsequently, the hot bubble cools and
diminishes very slowly.  See Figures~\ref{temperature}b, c, and d,
and \ref{density}b, c, and d.  
Between $1.6 \times 10^7$ and $1.7 \times 10^7$~yrs
it disappears entirely.  A more expanded description and additional
illustrations are provided in Paper I.

\subsection{Remnant A -- Overview to the X-Ray Emission}

X-ray emission rates are governed by multiple effects.  The important
emission processes 
(collisionally excited line, bremsstrahlung, two-photon, and recombination 
emission)
depend on $n^2$ and kinetic temperature.  Collisionally
excited line, two-photon, and recombination emission in the 
$1/4$ and $3/4$~keV bands additionally require that the metal
atoms be highly ionized.

	Typically, most of the soft X-ray power is due to line
emission from highly ionized, hot gas.
%Hot, diffuse plasmas produce $\sim100$ to $\sim300$~eV 
%soft X-rays mainly via line emission from highly ionized atoms.
In collisional equilibrium\footnote{The assumption of collisional
equilibrium is a commonly used tool for discussing X-ray production,
even in cases such as this, in which the ions are not in collisional 
equilibrium.}, 
the ions which emit strongly in the $\sim100$ to $\sim300$ eV 
energy range 
(useful for comparing with the ROSAT $1/4$~keV band countrate, the
Wisconsin All Sky Survey B and C bands countrates, or the 
lower part of ASCA's range)
%are {\bf{(fill in)}} which
are most abundant if the gas is $\sim10^6$~K.  
The ions that emit strongly in the  
$\sim400$ to $\sim1000$ eV energy range 
(useful for comparing with the ROSAT $3/4$~keV band countrate or the 
Wisconsin All Sky Survey M band countrate) are most abundant 
if the gas is $\sim5 \times 10^6$~K.
%  I took out this line because the SNR should not be very emissive above
%M band and answering Rob's question requires defining a band, defining
%dim, doing the calculations, etc. just to show that it isn't important.
%Except for a very short period {\bf{Rob asks how short}}, 
%the SNR is dim at higher energies.
%(see McCammon and Sanders Figure 2).
%{\bf{(check these)}}  {\bf{(fill in)}} are important ions for 
%this range.  
%In addition to line emission, the SNR also emits bremsstrahlung,
%two-photon, and recombination continuum emission.  These processes
%are more important in the SNR plasma than in collisional
%equilibrium plasma having the same color temperature.

Before the cool shell forms, the hot, comparatively dense gas behind
the shock front produces the vast majority of the SNR's X-ray emission.
%The gas's luminosity also depends on $n^2$.
%Because the density is much larger in the region just behind the 
%shock than it is in the bubble's interior, while the gas behind
%the shock front is very hot, it is responsible for
%the vast majority of the SNR's emission.
%the lionesses' share of the emission  
%{\bf{Rob and Steve don't like the lionesses' share business}} -- 
%until it cools.
After the cool shell forms, the region behind the shock is no
longer hot or highly ionized and so is no longer emissive. 
Thus, most of the emission derives from the 
remnant's ``hot'', highly ionized interior.  
As a result of
this shift, the spectrum and total flux level change
markedly.

%This section discusses the SNR's production of X-rays and provides the
%spectra and flux levels so that the results can be compared with past, 
%present, and
%future soft X-ray observations.  Finally, the conditions within the
%SNR are compared (and contrasted) with the conclusions that one would
%draw from hypothetical ROSAT observations of the SNR.

\subsection{Remnant A -- Ionization States}
%\noindent{\it{Ionization States}}

	The metal atoms in the hot bubble are seldom in
the collisional ionizational equilibrium with the gas temperature.
This challenge to the discussion of the plasma's ionization state
can be resolved by defining an ionization temperature, $T_i$.
%the ionization temperature and the gas
%temperature are often quite disparate.  
%For example, at the
%earliest epoch the center of the SNR is nearly $3 \times 10^7$~K,
%but as we saw in Figure~\ref{fig:oxyionfractions}
%few of the oxygen atoms are more highly ionized than \oxyfive\
%which in collisional equilibrium has a temperature of $3 \times 10^5$~K.
%{\bf{(if I break this into a separate paper, will have to drop that.)}}
If the out-of-equilibrium gas has the same distribution of ionization
states as collisional ionization equilibrium gas having a
kinetic temperature of $T_1$, then
the ionization temperature
%\footnote{The ionization temperature should
%not be confused with the kinetic temperature of the ions, which, in
%this simulation, is taken as equivalent to the kinetic energy of the
%electrons.} 
can be defined as the
equivalent equilibrium temperature, in this case $T_1$.  
Although the populations of ionization states in the SNR gas parcels
do not exactly match equilibrium populations and although different 
elements may indicate different $T_1$'s, 
it is still useful to compare ratios of prevalent ions
with the ratios found in collisional equilibrium 
in order to find approximate ionization 
temperatures, $T_i$'s.  
%Ionization temperatures were found from 
The ratios of Si$^{+8}$/Si$^{+9}$, Si$^{+7}$/Si$^{+8}$,
O$^{+6}$/O$^{+7}$, and O$^{+5}$/O$^{+6}$ ions were tested
as possible
indicators of the ionization temperature, with the
Si$^{+8}$/Si$^{+9}$ ratio chosen for presentation in
Figure~\ref{fig:iont}.
%The O$^{+5}$/O$^{+6}$ 
%ratio is a poor indicator because at many epochs the SNR interior
%contains little \oxyfive.  In contrast, O$^{+6}$ and O$^{+7}$ are
%relatively prevalent, making it a better indicator.  The
%O$^{+6}$/O$^{+7}$ ratio  The Si$^{+8}$/Si$^{+9}$ and 
%The ionization temperatures
%from the ratio of Si$^{+8}$/Si$^{+9}$ ions is presented in 
%Figure~\ref{fig:iont}.
%Although the results are not entirely independent of the choice of 
%indicators, $T_i$'s found from the Si$^{+8}$/Si$^{+9}$ ratio 
%are similar to those found from the O$^{+6}$/O$^{+7}$ ratio.
That figure presents the kinetic temperatures (the
temperatures corresponding to the random velocities of the atoms) 
and ionization temperatures
of the plasma for the five epochs prior to and just after the onset
of shell formation as well as for a selection of later epochs.

At $10^4$~years, the SNR is most dramatically 
underionized\footnote{In an underionized plasma, the ionization 
temperature is less than the kinetic temperature.}.  The
kinetic temperature, $T_k$, of the SNR is $\sim3 \times 10^7$~K, 
but the ionization temperature is only $\sim1 \times 10^6$~K.  
%The $T_k$ profile is flat due to thermal conduction and the
%$T_i$ profile is flat {\bf{(think about this more.)}}
%because the atoms in a given parcel underwent
%most of their ionization while the parcel was in the hot,
%relatively dense region just behind the shock and experienced
%few additional ionizations after the gas density dropped (when the
%shockfront moved on and left behind the parcel).
%Subsequently, $T_k$ drops rapidly and $T_i$ rises slowly.
The ionization temperature of the gas in the interior is rising
slowly because of the low collision rate in this low density, high
kinetic temperature gas.
The kinetic temperature in the interior
is dropping rapidly because the SNR's expansion spreads the interior's 
thermal energy across a rapidly increasing volume. 
Around $10^5$~yrs, the kinetic and ionization 
temperatures converge, largely due to the drop in the
kinetic temperature.
%not due to a 
%rise in the ionization temperature.  
%The ionization temperature of the hot bubble is fairly stable at 
By $2.5 \times 10^5$~yrs, $T_k$ in the interior is less than
$T_i$; thus the plasma has become overionized\footnote{In an
overionized plasma, the ionization temperature is greater than the
kinetic temperature.}.
The gas remains overionized for the rest of the SNR's life because
cooling (via emission, thermal conduction to more rapidly cooling gas,
and expansion) proceeds at a faster rate than recombination
(see Figure~\ref{fig:iont}b).

\subsection{Remnant A -- Spectra}
%\noindent{\it{Spectra}}

The nonequilibrium X-ray emission from 
the simulated remnant was calculated 
with the Raymond \& Smith spectral code (1977, 1993), using the
non-equilibrium ionic abundances calculated in the hydrodynamic simulation.
The spectra, 
%{\bf{(confirmed)}} 
shown in logarithmic form in 
%Figures~\ref{fig:spectralinear} and 
Figure~\ref{fig:spectralog}
%show the spectra in linear and log form for the SNR at
%five early epochs and four later ones.
includes line emission, a bremsstrahlung continuum, recombination edges, 
and two-photon continua.
%and a rounded continuum from $\sim$50~eV to $\sim$570~eV. 
The line emission from 50 to $\sim300$~eV is mainly due to highly ionized
N, O, Ne, Mg, Si, S, and Fe.
The line emission from $\sim300$~eV to $\sim600$~eV is mainly due to highly ionized
C, O, Ar, and Ca.
The lines between $\sim600$ and
$\sim900$~eV are mainly due to highly ionized N, O, and Fe.
The lines between $\sim900$ and 1150~eV are emitted mainly by highly ionized
Ne, Fe, and Ni.  
The O VII triplet ($\sim570$~eV) and the O VIII Lyman $\alpha$ line (653~eV) 
merit special attention.
They have been used as plasma diagnostics in the past
(Vedder, {\it{et al.}} 1986, Canizares (1990)) and may be
used in analyzing future high resolution data.
%The energy range spanned by the O VII ``triplet'' actually 
%covers four emission lines,
%the forbidden transition (1s$^2 -$ 1s2s ($^3S_1$)) at 561~eV, 
%the intercombination transition (1s$^2 -$ 1s2p ($^3P_1$)) at 569~eV, 
%a frequently overlooked average dielectronic recombination
%satellite transition (1s2nl $-$ 1s2pnl in which 
%$n =3$ dominates)
%%{\bf{($^3$S ?)}} 
%at 571~eV, and
%the resonance transition (1s$^2 -$ 1s2p ($^1P_1$)) at 574~eV.
%In an underionized plasma, the O~VII excited states are populated
%largely via collisional excitation, but also via
%inner shell ionizations of \oxyfive.  In this case, the resonance
%line is strongest.  In an overionized plasma, the O VII excited
%states are populated largely via recombinations from O$^{+7}$ to O$^{+6}$,
%as well as collisional excitations.  
%The $^3S_1$ and $^3P_1$ states have larger statistical weights than
%the $^1P_1$ state, increasing the probability that the electrons's
%route to the ground state will be via the forbidden or intercombination
%transitions.  In collisional equilibrium, collisional excitations of 
%O$^{+6}$, inner shell ionizations of \oxyfive, and recombinations of
%O$^{+7}$ each play a role.  
Furthermore, the spectra for 
$10^4$, $2.5 \times 10^4$, $5 \times 10^4$, and $10^5$~yrs shows a
rounded continuum running from $\sim$50~eV to $\sim$570~eV due to
a combination of two-photon continua emitted by O, C, and N.
The spectra for later epochs exhibit recombination edges.  For example,
the feature at 740~eV in the $10^6$~yr spectra is due to
O$^{+7}$ recombining to O$^{+6}$.

At $10^4$~yrs, the emitting plasma is exceptionally hot, but 
not yet highly ionized.  Some of the lines above about 700~eV are still
weak and the two photon continua below about 570~eV is very strong.
Between $10^4$ and $2.5 \times 10^4$~yrs, the ionization temperature 
rises by about $50\%$ (to about $1.5 \times 10^6$~K) and 
the kinetic temperature drops by a factor of three (to about $10^7$~K).  
%This has
%several ramifications, some of which are easily discerned in the 
%logarithmic spectral plot and some of which are not.
More lines populate
the high energy end, and 
the ratios of various emission lines shift.
%{\bf{(confirmed)}} 
%, for example
%the O VII resonance line becomes stronger than
%the nearby OVII forbidden line.
%
%and two-photon emission from neon 
%($\sim600$ and $\sim900$~eV) appears.
%Another oxygen two-photon continua develops between 570 
%and 650~eV, softening the edge of the first oxygen two-photon continuum.
%At $2.5 \times 10^4$~yrs the thermal bremsstrahlung emission 
%decreases with photon energy more steeply than before, 

In the ensuing 75,000~years, the kinetic temperature of the material
behind the shockfront decreases substantially, slightly softening 
the bremsstrahlung continuum.  The ionization temperature
begins to exceed the kinetic temperature and tiny recombination
edges begin to appear.  For example, the O$^{+7}$ to O$^{+6}$ 
recombination edge at 740~eV marginally appears in the $10^5$~yr spectrum.
The higher energy emission lines weaken.
%{\bf{(confirmed)}} 

%{\bf{(Need to improve this -- the strongest lines at all epochs are
%the ones below 100 eV.  Need to identify them and discuss them at
%relevant times.  Also need to put a set of labels on the plots -- could
%add another level (just make the plots $1/5$ taller) and put $|$ and
%labels for the lines around 70, 300, 350, 400, 600, 650.)  Do that
%for the first linear plot and put labels for the recombination edges 
%on the second linear plot.  Also point out that the ovii diminishes relative
%to the 300 and 350 lines by 100,000 years.}}

%Because the emission depends on $n^2$
%and the gas just behind the shock is much more dense than that in the 
%interior, this gas produces the lionesses' share of the emission
%-- until the cool shell forms.
Between $1.0$ and $2.5 \times 10^5$~yrs, 
%the gas behind the shock
%cools to $\stackrel{<}{\sim}10^5$~K and the shock slows to the point 
%where it is no longer able to thermalize gas to X-ray temperatures.  
%As a result, 
the gas within several parsecs of the shock front 
ceases to be X-ray luminous.  With only the SNR's tenuous interior 
contributing, the total X-ray luminosity wanes and the spectrum alters.
%Throughout the SNR,
%the kinetic temperature drops below the ionization temperature.  
Lower energy emission lines dominate the spectrum.
Recombination edges become more prominent, while line and two-photon
emission diminish.
%{\bf{(confirmed)}} 

The remnant's physical characteristics continue to evolve until an age of
about one million years. Then for several million years afterwards, 
the kinetic and ionization temperatures remain nearly constant.
%By $5 \times 10^5$~yrs, the interior cools 
%to $~\sim10^6$~K and by $10^6$~yrs it had cooled to 
%For the ensuing several million years, the
%temperature remains nearly constant.  
%The ionization temperature, which is nearly
%twice the kinetic temperature, is also nearly constant with respect to time. 
Similarly, the spectra radically evolve between $2.5 \times 10^4$ and
$10^6$~yrs, then remain fairly unchanged.   
Compared with the spectra from earlier epochs, the spectra for
1, 5, and $10 \times 10^6$~yrs (Figure~\ref{fig:spectralog}) have 
reduced strengths in the
higher photon energy emission lines,
increased prominence in the recombination edges and 
steepened slopes in the recombination continua.
%{\bf{(confirmed)}}  

At $1.5 \times 10^7$~yrs, the kinetic temperature 
(only $\sim10^5$~K and less)
%The kinetic temperature 
begins to rapidly drop and the spectral characteristics change again.  
Now, the extremely steep bremsstrahlung component is much dimmer than
those from the other processes.  
An ensemble of two-photon emission curves defines
the shape of the continuum between 250 and 900~eV.  A series of
recombination edges punctuate the spectrum at 54, 126 and 387 eV
and a relatively sparse collection of emission lines rises above the
continua.  (Note that for the low temperatures of this epoch, 
the Raymond and Smith spectral code terminated 
some predictions below 920 eV.)

%Si$^{+7}$ is presented in Figure~\ref{fig:siionfractions}.
%{\bf{(make figure)}}  This ion emits at several frequencies, with
%especially strong emission at 195 and 203 eV (Raymond \& Smith, 1977).
%{\bf{(maybe switch to the Sulfur $+8$)}}.
%For the higher energy range, 
%O$^{+6}$ and O$^{+7}$ were chosen.
%The O$^{+6}$ has a strong complex of lines around
%560 to 570 eV, and the O$^{+7}$ has a strong transition
%at 666 eV.
%%a forbidden line at 561 eV, an intercombination 
%%line at 569 eV, a satellite line at 571 eV, and a resonance
%%line at 574 eV.  
%One or both of these lines have been observed in various 
%pointings toward the northern sky
%(Inoue {\it{et al}}., (1979 and
%1980, 
%%with 200 eV FWHM resolution proportional
%%counters, 
%Schnopper {\it{et al.}} 1982 Rocchia {\it{et al.}}
%1984 {\bf{(check those, came from McCammon and Sanders)}}, 
%Gendreau {\it{et al.}} 1995, and the XQC group (1997 - in preparation?)).
%The ionization fraction plots are 
%presented in Figure~\ref{fig:oxy78ionfractions}

\subsection{Remnant A -- Total Luminosity and Spatial Appearance}
%\noindent{\it{Total Luminosity and Spatial Appearance}}

The SNR's luminosity and spatial appearance are most easily presented in
terms of broad band countrates.  For this case, the
ROSAT PSPC $1/4$ and $3/4$~eV bands are used.
The ROSAT PSPC $1/4$~keV band is composed of the R1 band 
($\sim110$ to 284 eV) and the 
R2 band ($\sim$140 to 284 eV), while the ROSAT PSPC 
$3/4$~keV band is composed of the R4 band ($\sim$440 to $\sim$1010 eV)
and the R5 band ($\sim560$ to $\sim1210$eV).
%band countrates for 
Figure~\ref{bandfunctions} depicts the band response functions.
The R2 band is harder than
%more sensitive to 100 to 284 eV photons than is 
the R1 band, although both bands have the same upper boundary at 284 eV.  
The R2 band also has some sensitivity above 500~eV.
%(Snowden 
%{\it{et al.}} 1994). {\bf{Steve said that he has the post script file
%and that I should add the figure for the response functions.}}
The simulated spectra discussed
in the previous subsection were convolved 
with the ROSAT
PSPC response matrix, effective area, window transmission and gas
transmission coefficients (Briel {\it{et al.}} 1996) to determine the
following results.

%	The densest part of the SNR is the region behind the shock 
%front.  
During the first few epochs, the hot, relatively dense gas just behind
the shock produces most of the soft X-rays, causing the remnant
to appear luminous (Figure~\ref{fig:totalxray}) and edge brightened
(Figures~\ref{fig:R1flux} and \ref{fig:R4flux}) 
%{\bf{(confirmed)}}.  
Between $5 \times 10^4$ and $10^5$~yrs, the spectra softens. 
Consequently, the
SNR dims and loses its edge brightened appearance in the $3/4$~keV band.  
Between $10^5$ and $2.5 \times 10^5$~yrs,
the densest part of the SNR cools to less than $10^5$~K;
the ``cool shell'' forms and the
shock front weakens substantially.  
The region behind the shock now dims in the $1/4$~keV band.
With only the SNR interior providing X-rays, the SNR ceases to appear
edge brightened or bright.  During the course of a couple million
years, the total $1/4$~keV luminosity diminishes by
a factor of $\sim 100$. 
%{\bf{(confirmed)}}
It remains near this level until around $1.5 \times 10^7$~yrs
when the ancient, relatively cool,
relatively dense SNR begins a phase of rapid decline.
Integrating the $1/4$~keV luminosity
with respect to time yields 
$5.8 \times 10^{59}$ counts~cm$^{2}$ 
%{\bf{(confirmed)}}.  
Integrating the $3/4$~keV luminosity
with respect to time yields
$1.2 \times 10^{58}$ counts~cm$^{2}$~yr~s$^{-1}$.

\subsection{Remnant A -- The \carthree, \nitfour, and \oxyfive\ Content}

%Some potential halo models can be tested with 
%data from a number of wavelength ranges.
%The halo SNRs
%are no exception.  They produce copious quantities of 
%\carthree, \nitfour, and \oxyfive\ which are detected in the ultraviolet
%part of the spectrum.  
%While Paper I elaborately describes
%the high-stage ions in Remnant A, this section summarizes the
%results for Remnants A and B.  

Paper I provided a detailed study of the \carthree, \nitfour, and 
\oxyfive\ ions in Remnant A.  This subsection summarizes the important
results.
While the SNR is young, (for example, at $10^4$~yrs) its
%remnant is filled with hot ($>10^7$~K) gas, but its ions are far
%from equilibrium. The 
atoms are ionizing up through the
\carthree, \nitfour, and \oxyfive\ states.  
As a result, very large column 
densities of UV ions exist in the young remnant, however
their bulk velocities and thermal broadening can be enormous.  
By $2.5 \times 10^4$~yrs, the interior gas has ionized beyond
these states and the \carthree, \nitfour, and \oxyfive\ exist
only in the recently heated, underionized gas just behind the shock
front.
%At this age, the remnant is also very luminous
%in X-rays {\bf{(put this in terms of observations)}}.
%The soft X-ray luminosities peak within the first 100,000
%years and drop fairly rapidly afterward.  The UV ion 
%content peaks much later, around 3 $\times 10^6$~years, drops off 
%much more slowly, and is significant to $\sim2 \times 10^7$~years.
As the shell forms, the shock front becomes too weak
to ionize the swept up ambient gas up to the \carthree,
\nitfour, and \oxyfive\ states.  Henceforth, these ions only derive
from cooling, recombining gas.  Highly ionized oxygen can recombine to
O$^{+4}$ before the nitrogen recombines to N$^{+3}$ or the carbon 
recombines to C$^{+2}$ and the gas cools before all of the carbon
has recombined to the C$^{+2}$ level.  As a result, 
the \carthree\ extends out to greater radii in the cooling gas
than does the \nitfour\ or \oxyfive, and at very late times it resides
in cooled gas. In Remnant A,  
some \carthree\ remains a million years after the SNR cools.
The column densities are only weakly dependent on the age of the 
SNR, particularly after $5 \times 10^4$~yrs.  
%The time and
%impact parameter averaged column densities for a single SNR are
%$5 \times 10^{13}$ \oxyfive cm$^{-2}$, 
%$5 \times 10^{12}$ \nitfour cm$^{-2}$, 
%$1 \times 10^{13}$ \carthree cm$^{-2}$.

The time integrals of the number of high-stage ions contained by 
Remnant A are
$7.8 \times 10^{69}$ \oxyfive seconds, $7.2 \times 10^{68}$ \nitfour seconds,
and $1.6 \times 10^{69}$ \carthree seconds.  Dividing by the time integral
of the area covered by the \carthree, \nitfour, or \oxyfive\
ions gives the time and 
impact parameter averaged column densities for a sightline which intersects
the high stage ions\footnote{A subtilty is that the
\carthree\ covers $15\%$ more area than the \oxyfive.}: 
$5.2 \times 10^{13}$ \oxyfive cm$^{-2}$, 
$4.7 \times 10^{12}$ \nitfour cm$^{-2}$, and $9.8 \times 10^{12}$ 
\carthree cm$^{-2}$.
%The time integrated luminosities for Remnant A are
%$1.4 \times 10^{49}$ ergs in the \oxyfive\ doublet,
%$1.9 \times 10^{48}$ ergs in the \nitfour\ doublet, and
%$4.5 \times 10^{48}$ ergs in the \carthree\ doublet 
%{\bf{(Sept 1998 need to confirm)}}.  
%%This SNR is very
%%bright while it is very young, but dims substantially after the cool
%%shell forms.  For this reason, the time averaged fluxes would not be 
%%be meaningful.

\section{Remnant B:
$n_o = 0.01$~cm$^{-3}$, $P_{nt} = 7.2 \times 10^3$, and
$E_o = 0.5 \times 10^{51}$~ergs}

Remnant B has four times the ambient nonthermal pressure as Remnant A
and half of the explosion energy of Remnant C.
By comparing the structures, evolutions, spectra, 
and luminosities of these remnants
we can see the effects of varying these parameters.

\subsection{Remnant B -- Structure}

Figures~\ref{temperature5uG}, \ref{density5uG}, 
and \ref{pressure5uG} depict Remnant B's kinetic temperature, density of
atoms, and thermal and total pressure as a function of radius
for a variety of ages.  In the original simulation
the shockfront traveled to the edge of the grid, reflected, and
traveled back towards the hot bubble, colliding with the bubble
between 1.1 and $1.2 \times 10^7$~yrs which is
just before the end of the SNR's life.  Thus the post 
$1.1 \times 10^7$~yrs results have been replaced with those from
an additional, simulation performed with half the spatial resolution.
The SNR has completely disappeared by $1.3 \times 10^7$~yrs.

Before their cool shells form, this remnant and Remnant A are
similar in size, temperature, density, and thermal and total pressure.
The early evolution has been little affected by quadrupling the
ambient nonthermal pressure.
Sometime after their cool shells begin to form 
(which occurs only slightly earlier
in Remnant B than in Remnant A), however, each bubble's thermal pressure
comes to approximate the total ambient pressure.  Thus, after the
shell forms, Remnant B's bubble evolves to be of higher thermal pressure, 
hotter, and denser than Remnant A.  Consequently
Remnant B radiates away its energy on a shorter timescale
and so contains a warm or hot bubble for only $\sim 1.2 \times 10^7$~yrs, 
whereas Remnant A's bubble was warm or hot for $\sim 1.6 \times 10^7$~yrs, 

%At $1.1 \times 10^7$~yrs, the hot bubble of Remnant B is very near 
%the end of its life: the maximum temperature in the interior is only 
%$\sim10^5$~K and the radial extent is only $\sim40$~pc.  
%For comparison, remnant
%A does not mature to the stage of having this interior temperature and 
%radial extent until
%$1.5 \times 10^7$~yrs.  
%Remnant A disappears before another 2 million years have passed,
%during which time it is a very dim X-ray emitter.
%Remnant B would probably do the same, but in the simulation, the
%shockfront has traveled to the edge of the grid, reflected, and
%returned to impact with the hot bubble between 
%1.1 and $1.2 \times 10^7$~yrs.  The impact compresses the hot bubble,
%causing the elevation in temperature, density, and pressure seen at
%$1.2 \times 10^7$~yrs and causing the disturbances seen in the
%later curves.  
%Additional simulations showed that collisions with reflected shocks
%caused the remnants to emit more readily for a short time, does
%not significantly alter the time integrated results.  
%Considering that the the shock collision
%occurred during the final, dim megayears of the SNR's life,
%its response to the shock collision
%should have little effect on the overall conclusions drawn from the
%simulations.

\subsection{Remnant B -- Spectra, Total Luminosity, and Spatial Appearance}

During Remnant B's energy conserving phase, its 
spectra (Figure~\ref{fig:spectralogB})  
%{\bf{(confirmed)}} 
is nearly identical to that of Remnant A.
Small differences begin to appear around 250,000~yrs and
are obvious after about a million years.
Many of the emission lines below 400~eV grow to be much stronger 
(even an order of
magnitude stronger) than the lines in Remnant A and
until Remnant B is on the brink of death, its recombination
spectra are not as strongly sloped as those of Remnant A at the same ages.
%The time integrated spectra for Remnants
%A, B, and C are plotted in Figure~\ref{fig:cumulative}.  
%{\bf{(In general, they are very similar.}}

Figure~\ref{luminosityB} presents
Remnant B's luminosity as a function of time
%{\bf{(confirmed)}}. 
For the first $2.5 \times 10^5$~yrs
the luminosities in the ROSAT PSPC $1/4$ and $3/4$~keV bands
nearly trace those of Remnant A.  Afterward, both remnants dim,
but because of its larger temperature, density, and thermal pressure, 
Remnant B dims less.  
%  the halo29 + halo30 combination gives the same rates
Integrating the
$1/4$~keV luminosity 
with respect to time, yields 
$1.4 \times 10^{60}$ counts~cm$^{2}$.
%{\bf{(confirmed)}}.  
The integrated $3/4$~keV luminosity 
luminosity is $1.1 \times 10^{58}$ counts~cm$^{2}$.
%{\bf{(confirmed)}}. 
Thus, during
its lifetime, Remnant B produces over twice as many $1/4$~keV photons
as Remnant A, and, unlike Remnant A, produces most of its
$1/4$~keV photons after the cool shell forms.
%{\bf{(confirmed)}}.  
Its production of $3/4$~keV photons is no higher than that of Remnant A.
%{\bf{(confirmed)}}
%Remnant B emits $2\%$ of its total radiated energy in the form of
%100~eV and hotter photons {\bf{(Sept 1998 need to confirm)}}.
%%  To more significant digits, the number is 1.7%

Remnant B's spatial appearance can be discussed in terms of its
surface brightness as a function of impact parameter.  Like Remnant
A, Remnant B initially appears edge brightened and later evolves to
appear centrally filled.
(see Figure~\ref{fig:halo29R1flux}
%\ref{fig:halo29R2flux}, 
and \ref{fig:halo29R4flux}).
%The evolution matches that of Remnant A, except that an old
%(post cool shell formation)
During its youth, Remnant B has a similar luminosity and spatial appearance
as Remnant A.
During its centrally filled phase, however, the ancient
Remnant B's $1/4$~keV flux is a couple hundred $\times 10^{-6}$
counts s$^{-1}$ arcmin$^{-2}$, which is
about ten times greater than that of Remnant A and 
%During this time, Remnant B's $1/4$~keV
%surface brightness is
%about $200 \times 10^{-6}$ counts s$^{-1}$ arcmin$^{-2}$ 
%{\bf{(Sept 1998 need to confirm)}}, making 
bright enough to be observable to an analysis like that done by
Snowden {\it{et al.}} (1998).

\subsection{Remnant B -- The \carthree, \nitfour, and \oxyfive\ Content}

The time integrals of the number of high-stage ions contained in
Remnant B are $6.0 \times 10^{69}$ \oxyfive seconds,
$4.6 \times 10^{68}$ \nitfour seconds, and $1.1 \times 10^{69}$ 
\carthree seconds.  After the cool shell forms, the number of ions
contained by the remnant is a slowly varying function of time and
impact parameter.  As a result,
the age and impact parameter averaged column densities (found by
dividing by the time integrals of the number of ions by the
time integrals of the areas\footnote{The
average area covered by \carthree\ ions is greater than that
covered by \nitfour\ or \oxyfive\ ions.})
is a reasonable estimate of the column
density expected for a sightline which traverses the high-stage ions
in and around the hot bubble.  These values are:
$7.8 \times 10^{13}$ \oxyfive cm$^{-2}$,
$5.7 \times 10^{12}$ \nitfour cm$^{-2}$, and 
$1.2 \times 10^{13}$ \carthree cm$^{-2}$.
%Remnant B's time integrated emission 
%in the \oxyfive\ doublet is $6.88 \times 10^{49}$ ergs, its emission 
%in the \nitfour\ doublet is $2.32 \times 10^{48}$ ergs, and its emission
%in the \carthree\ doublet is $4.39 \times 10^{48}$ ergs.
%The \oxyfive\ emission is about 4 times greater than that of
%Remnant A.  In contrast, the \carthree\ emission is about equal to that of
%Remnant A {\bf{(Sept 1998 need to confirm)}}.

\section{Remnant C:
$n_o = 0.01$~cm$^{-3}$, $P_{nt} = 7.2 \times 10^3$, and
$E_o = 1.0 \times 10^{51}$~ergs}

%With twice the explosion energy of the previous simulation, Remnant C
%is a larger, hotter, longer lived, and brighter version of Remnant B.

\subsection{Remnant C -- Structure}

The kinetic temperature, density of atoms, and thermal and total
pressure are plotted in Figures~\ref{fig:temperatureC}, \ref{fig:densityC}, 
and \ref{fig:pressureC}.  
Remnant C is hotter, larger, more diffuse,
generally more pressurized, and longer lived.
than its lower explosion energy cohort, Remnant B. 

\subsection{Remnant C -- Spectra, Total Luminosity, and Spatial Appearance}

Figure~\ref{fig:spectralogC}
%{\bf{(confirmed)}}
depicts Remnant C's spectra.  For the first 5 million years,
the spectral features nearly trace those of Remnant B, except that 
Remnant C is brighter.  
%By 10 million years, their spectra have begun to differ noticeably.
%This is because Remnant A is still hot, but 
%Remnant B is in its final few million years of life and so is much cooler.
Correspondingly,
Remnant C's
luminosity as a function of time (Figure~\ref{fig:luminosityC})
follows the pattern set by
Remnant B, but with a greater magnitude.
%but is greater by a factor of approximately 2 $\frac{1}{2}$
%{\bf{(confirmed)}}.
The time integrated 
ROSAT PSPC $1/4$~keV luminosity is
$3.5 \times 10^{60}$ counts cm$^{2}$, which is 
2.3 times that of Remnant B.
%{\bf{(confirmed)}}.
Over $80\%$ of this
emission is produced after the cool shell forms.
The time integrated
ROSAT PSPC $3/4$~keV luminosity
is $2.8 \times 10^{58}$ counts cm$^{2}$, which is 
2.5 times that of Remnant B.
%{\bf{(confirmed)}}.
Only about $25\%$ of the $3/4$~keV 
emission is produced after the cool shell forms.  
%For Remnant C, $2\%$ of the energy is emitted in 100~eV and hotter photons
%{\bf{(Sept 1998 need to confirm)}}.
%%  to more significant digits, the number is 2.4%

The spatial appearance (Figures~\ref{fig:halo31R1flux} and
%\ref{fig:halo31R2flux}, and 
\ref{fig:halo31R4flux})
is edge brightened during the remnant's energy conserving
phase and centrally filled during the post shell formation phase.
The progression is very similar to that of Remnant B.  
%One notable
%difference is that
%the ancient Remnant C's flux has somewhat lower ratios of 
%R4 to R2 counts and R2 to R1 counts {\bf{(Sept 1998 need to confirm)}}. 

\subsection{Remnant C -- The \carthree, \nitfour, and \oxyfive\ Content}

The time integrals of the number of high-stage ions contained in
Remnant B are $1.2 \times 10^{70}$ \oxyfive\ seconds, 
$8.9 \times 10^{68}$ \nitfour\ seconds, and
$2.4 \times 10^{69}$ \carthree\ seconds.
Dividing these time integrals by the time integrated area covered by
each of the respective ions yields the
age and impact parameter averaged column densities.  They are
$8.7 \times 10^{13}$ \oxyfive\ cm$^{-2}$, 
$6.0 \times 10^{12}$ \nitfour\ cm$^{-2}$, and
$1.5 \times 10^{13}$ \carthree\ cm$^{-2}$.
%The time integrated emission in the \oxyfive, \nitfour, and
%\carthree\ doublets are 
%$4.5 \times 10^{49}$ ergs, $4.4 \times 10^{48}$ ergs, and
%$8.3 \times 10^{48}$ ergs, respectively {\bf{(Sept 1998 need to confirm)}}. 

\section{The Observer's Perception of the Simulated Data:
Color Temperature, Electron Density, Thermal Pressure, and
Oxygen line diagnostics}

Oft-used observational analysis procedures include determining
an object's spectrum's color temperature, electron density, and
thermal pressure.  In addition, with very high spectral resolution
data, the ratios of the O~VII emission line fluxes
and the ratio of the O~VII to O~VIII line fluxes have been 
used as diagnostics of the plasma's temperature and ionization history.
In this section, the simulated data is subjected to the identical
treatment in order to determine how the ``observed quantities'' map 
to the original physical properties of the emitting gas and to 
provide observational signatures of a SNR evolving in
a diffuse, pressure dominated region.
Remnant A is used as the test case.  Considering that Remnants B and
C are hotter, they may have higher color temperatures.

\subsection{Color Temperature}

%{\bf{Rob says that he doesn't understand this -- ``Don't you want
%to fold your spectra (nei) through the ROSAT bands, obtain
%a ratio, and compare these with $T_c$.}}
One wonders how hot one of the simulated SNRs would appear to an
astronomer using only ROSAT data.  In order to answer this question,
a temperature yardstick must be created.  (As is commonly done
in analyses of observational data, the yardstick is calibrated
as if the observed spectrum derives from ions which are in collisional
equilibrium.)  While the assumption may be far from true, the 
{\underline{apparent}} temperatures are still useful
as descriptors of the measured spectra.  
Following the example of Snowden
{\it{et al.}} (1998) a temperature measure (called the color
temperature, $T_c$) was crafted by
calculating the ROSAT PSPC R1/R2 and R2/R4 band ratios for
equilibrium spectra simulated with the Raymond \& Smith code (1977, 1993),
Grevesse and Anders (1989) abundances, and various assumed temperatures.
The R1/R2 ratio is steep and single valued
(making a good yardstick) between $T_{cR1R2} = 10^5$ and $10^6$~K.
%{\bf{(confirmed)}}.  
The R2/R4 ratio is single valued 
below $T_{cR2R4} = 2.5 \times 10^6$~K.
%{\bf{(confirmed)}}.  
%The combination
%of ratios provides a uniquely-valued yardsticks for 
%higher temperatures.  There is
%little need to use such a device, because $T_{cR2R4}$
%rarely approaches $2.5 \times 10^6$~K in this remnant.  

See Figure~\ref{fig:colort} for the color temperature versus impact 
parameter.  Interestingly, 
even when both color temperatures are within their single-valued ranges, 
the R1/R2 color temperature is generally less than the
R2/R4 color temperature.
%$T_{cR1R2} < T_{cR2R4}$.

Some evolution of the color temperatures does occur, 
but the most striking characteristic of the SNR
is how small that variation is.
At $10^4$~yrs, the spectrum from a 
pointing toward the center of the remnant 
has a R1/R2 color temperature of about $10^6$~K and a 
R2/R4 color temperature of about
$2.0 \times 10^6$~K, while the spectrum from 
a pointing toward the edge of the remnant has slightly lower
color temperatures.   
Between $10^4$ and $2.5 \times 10^4$~yrs, $T_{cR1R2}$ as a function of
fractional impact parameter
increases slightly while $T_{cR1R2}$ as a function of fractional
impact parameter remains approximately constant.  
From $2.5 \times 10^4$ to $10^5$~yrs, $T_{cR1R2}$ and $T_{cR2R4}$
drop slowly.
Over the remainder of Remnant A's lifetime, the 
the R1/R2 color temperature drops to several hundred thousand degrees,
the R2/R4 color temperature hovers just above a million degrees, and
the profiles develop and upward curve.
%{\bf{(confirmed)}}.
This exercise is telling us that the observationally determined
color temperature may not directly reveal 
either the kinetic or ionization temperatures, yet it may be possible to use
simulations to interpret the observationally determined color temperature.
A note of clarification:  As x-ray photons transit interstellar
material, the lower energy photons are preferentially absorbed, making
the spectrum appear 'harder', as if it derives from hotter plasma.  Thus,
in order to determine the true spectra, the observed spectra must be
'de-absorbed', as is done in Snowden, {\it{et al.}} (1998)'s
determination of the halo color temperature.  In contrast, de-absorption
is hardly an issue in analyzing the Local Bubble spectrum because that
light transits a negligible absorption column density before reaching the
solar system.

%The SNR is brightest when it is very young, but spends most
%of its lifetime in its dim, older stage.   
%The countrates were integrated in order to find the color 
%temperature corresponding to a background flux emitted by an 
%ensemble of halo SNRs.
%The time and
%space averaged R1/R2 ratio of 1.7 leads to $T_c = 7 \times 10^5$~K
%while the R2/R4 ratio of 20 leads to $T_c = 1.6 \times 10^6$~K.
%{\bf{(compare the color temperature vs ratio with Steve!)}}

%The color temperatures may be useful as a means 
%for comparing the simulations with observations, but
%any comparisons should be consider the following.  
%The R1/R2 and the R2/R4
%ratios give different color temperatures, the
%color temperatures do not consistently correlate with either
%the kinetic or the ionization temperatures, and
%the assumed abundances, choice of spectral code, and the assumption of 
%electron-ion equilibrium may affect the calculated
%color temperatures {\bf{(Sept 1998 need to confirm)}}.\\

\subsection{Electron Density and Thermal Pressure}

In this subsection, the electron density ($n_e$) and thermal
pressure ($P_{th}$) are calculated
using an observational-type analysis of the
simulated emission and compared with the $n_e$ and $P_{th}$
taken directly from the simulation.  Using low spectral resolution
detectors, such as those on ROSAT, the
observationally determined variables are often inconsistent with
the true values, suggesting that further progress in connecting the
broad band data with the source phenomena may be possible via mappings
such as the following.
%can be made via observational-type analyses of simulations of possible
%sources.

The standard analysis finds $n_e$ from the emission measure
($n_e^2(l) dl$), where $l$ is the depth of the
luminous region and is generally taken 
as equal to its width or a simple function of the width).
The emission measure 
is calculated from the observed countrate as follows.
%using the assumption that the emission measure equals
%the observed countrate divided by
%a calculated countrate per emission measure.  
%where the countrate per
%emission measure is calculated as follows. 
The color temperature (found from the observations) is taken as an
approximation of the kinetic temperature (this is known
to be a false approximation), the gas is assumed to be in ionizational
equilibrium (also generally false), using these temperatures
as inputs and using a given emission measure, the high energy
resolution spectrum is then calculated with a spectral code 
(such as the Raymond and
Smith code) and folded through the detector response
functions.  That act yields a countrate per emission measure which is
then compared with the observed countrate in order to determine the
emission measure.

A hypothetical 
pointing toward the center of the $10^4$~year old simulated remnant
(Remnant A) has an R2 countrate of 
$1.1 \times 10^{-3}$ counts s$^{-1}$ arcmin$^{-2}$,
an R1/R2 color temperature of $1.2 \times 10^6$~K, and a depth of
60~pc.  Assuming that the color temperature approximates
the kinetic and ionization temperatures and using Grevesse and
Anders (1989) abundances, spectral calculations made with the
Raymond and Smith (1977, 1993) spectral code
yield an R2 countrate per emission measure of  
$2.3 \times 10^{-20}$ counts s$^{-1}$ arcmin$^{-2}$ cm$^{5}$.
%The width (and depth) of the luminous region is 60~pc.
The electron density calculated from the unrounded values is 
%For this the countrate from the
%center of the halo SNR is $1.2 \times 10^{-3}$ R2 counts s$^{-1}$ 
%arcmin$^{-2}$.  At the R1/R2 color temperature of $1.1 \times 10^6$~K,
%%an emission measure of 1 cm$^{-5}$ produces
%%$3.5 \times 10^{-20}$ R2 counts s$^{-1}$ arcmin$^{-2}$.  Thus, 
%%$\int n_e^2(l) dl = 
%the emission measure is $3.4 \times 10^{16}$~cm$^{-5}$. Taking 
%$\int n_e^2(l) dl \sim n_e^2 \times l$ and 
%the diameter (60pc) as the depth, yields an electron density of 
0.016~cm $^{-3}$.  Assuming full ionization of the hydrogen and helium
and using $T_c$ as $T_k$ yields $P_{th}/k = 3.8 \times 10^4$~K cm$^{-3}$.
How do these values compare with the emission measure, temperature,
and thermal pressure found directly from the hydrocode?
The observational-type analysis implies an $\int n_e^2(l) dl$ that is roughly 
3 times higher than the $\int n_e^2(l) dl$ calculated directly 
from the hydrocode;  the color temperature
is two orders of magnitude below the kinetic temperature, and the
estimated thermal pressure is one seventh of that from the hydrocode.
By $10^5$~yrs, the values have come into better agreement.  
By $10^6$~yrs, the $\int n_e^2(l) dl$ calculated by this method is 
an order of magnitude less than the $\int n_e^2(l) dl$ calculated
directly from the hydrocode, the color temperature is about
$3/2$ the kinetic temperature, and the expected thermal
pressure is less than the thermal pressure
calculated directly from the hydrocode.\\
%{\bf{(Sept 1998 need to
%confirm)}}.\\

%	The OVII-OVIII and line/continuum results
%	are affected by the assumption
%	of electron-ion equilibrium.

\subsection{O VII and O VIII Emission Lines}

We are entering an era in which
very high resolution X-ray spectrometers will record
the signals of individual lines and complexes.  
The spectra should be complex and difficult to fit with 
simple models.  One approach to the complexity is
to extract physical information from specific emission lines.  
The O~VIII Lyman $\alpha$ line at 653 eV and 
the O~VII ``triplet'' around 570 eV are 
good candidates for examination because
these lines provide useful measures of the temperature and ionization
state, are strong, are not closely clustered with other strong lines, 
and because model estimates depend on relatively well known atomic
constants.

%{\bf{(check for repetition with the spectral section)}}
The energy range spanned by the O VII ``triplet'' actually 
contains several emission lines.  There are
forbidden (1s1s $-$ 1s2s ($^3S_1$)),
% at 561~eV, 
intercombination (1s1s $-$ 1s2p ($^3P_1$) and 1s1s $-$ 1s2p ($^1P_1$)),
%at 569~eV, 
%the frequently overlooked 
%satellite transition (1s2nl$^2 -$ 1s2pnl) at 571~eV, and
resonance (1s1s $-$ 1s2p ($^1P_1$)),
%at 574~eV.
and dielectronic recombination satellite 
(1s2nl$^2 -$ 1s2pnl) lines, as well as a 
line due to innershell excitation while the
oxygen is in its \oxyfive\ state.
The Raymond and Smith code combines the strengths of the
two intercombination lines and reports them as a single entry
at 569 eV.  Currently the code does not consider innershell excitation
of oxygen in its \oxyfive\ state.

	Thus, the spectra presented here include four lines,
the forbidden (561 eV), merged intercombination (569 eV), 
satellite (571 eV), and resonance (574 eV).  The forbidden,
intercombination, and resonance transitions can result from
collisions of O$^{+6}$ or recombinations from O$^{+7}$ to O$^{+6}$.
In contrast, the dielectronic recombination satellite transition can only
result from recombinations from O$^{+7}$ to O$^{+6}$.
In collisional equilibrium, collisional excitations of 
O$^{+6}$, inner shell ionizations of \oxyfive, and recombinations of
O$^{+7}$ each play a role.  
In an underionized plasma, the O$^{+6}$ excited states are populated
largely via collisional excitation, but also via
inner shell ionizations of \oxyfive.  In this case, the resonance
line is strongest.  In an overionized plasma, the O VII excited
states are populated largely via recombinations from O$^{+7}$ to O$^{+6}$,
as well as collisional excitations.  
The $^3S_1$ and $^3P_1$ states have larger statistical weights than
the $^1P_1$ state, increasing the probability that the electrons's
route to the ground state will be via the forbidden or intercombination
transitions.  

Figure~\ref{fig:ovii} presents the 
line luminosities of the four O VII lines and the O VIII Lyman alpha
(653 eV) line.
There is a tradition of plotting the
ratio of the O VII forbidden to resonance line luminosities against the
the ratio of the O VIII to O VII lines for underionized plasmas, 
(Vedder {\it{et al.}} (1986), Canizares (1990), and 
Sanders {\it{et al.}} (1997))
The ratio of the O VII lines is thought to provide an indicator of the
ionization state of the plasma (though the temperature also plays a role).
The ratio of the O VIII to O VII luminosities is thought to 
provide an indicator of the temperature, with the provisos that
the ionization time plays a role 
in underionized plasma,  under and overionized plasmas produce
different O VIII to O VII ratios, and 
above $T {\sim} 2 \times 10^6$~K, 
the choice of equilibrium, isobaric or isochoric cooling
plays a role for cooling plasma.
%, as can be seen
%by examining the steady state and isochoric cooling 
%plots of Schmutzler and Tscharnuter (1993) and from equilibrium and
%isobaric cooling simulations done using the
%Gaetz, Edgar, and Chevalier (1988) transition rate table).
One subtlety is that depending on the spectral resolution, 
the intercombination and satellite lines may or may not be 
resolvable from the resonance line. 
Figure~\ref{fig:oviioviii} is a variation on the standard figure, having
the ratio of the O~VII forbidden to resonance plus satellite line 
luminosities plotted against the ratio of the O~VIII to O~VII resonance 
plus satellite line luminosities.  The satellite line is much
weaker than the resonance line; excluding it does not alter the character of
the curve.  The figure also includes the
ratios for a plasma in ionizational equilibrium.  Underionized plasmas
correspond to the region below this curve and overionized plasmas correspond
to the region above the curve.

%As can be seen in Figure~\ref{fig:ovii}
%the intercombination line can be as strong
%as the resonance line.
%Sanders {\it{et al.}} (1997) have addressed this issue to some degree by
%using the ratio of O VII forbidden line to the sum of the resonance
%and intercombination line and the ratio of the O~VIII Lyman $\alpha$ 
%line to
%the sum of the O~VII resonance and intercombination line.
%Their Figure 2 includes the trajectories for an underionized plasma
%as well as the trajectory for Remnant A presented in this project.
%Figures~\ref{fig:oviioviii}b and c include these additional lines
%{\bf{(Sept 1998 need to confirm)}}.

%Figure~\ref{fig:ovii} presents the SNR's luminosity 
%in the O VII 561~eV, 569~eV, 571~eV, and 574~eV lines, as well
%as the O VIII 653~eV line as a function of time.

Aside from the first datapoint ($10^4$ yrs), the OVII - OVIII 
signature of the young SNR accurately describes it as
an underionized plasma having a low ionization parameter
($nt$, where t is the age) and a fairly high 
thermal temperature.
A quasi ionizational equilibrium is reached when the kinetic temperature
drops down to the ionization temperature at 
$\sim1 \times 10^5$~yrs (epoch 4).  Afterwards, the gas is
recombining, and predictably, the OVII forbidden to resonance 
line ratio rises.  It peaks at $10^6$ yrs (epoch 7) before it and
the O~VIII to O~VII ratio begin their slow descents.  

\section{Caveats}

Paper I discussed a long list of assumptions and approximations
pertaining to the modeling
and interpretation of the \carthree, \nitfour, and \oxyfive\ 
calculations.  Additional considerations, which more strongly affect the
X-ray emission are presented here along with the most pertinent
caveats from Paper I.

In the hydrocode, the electron and ion kinetic temperatures 
are set equal to each other, although during the earliest part of the
evolution, the
electrons should be cooler than the ions.  To find an upper limit on
the timescale for the electrons to come into equilibrium with the
ions, they will be assumed to 
equilibrate solely via Coulomb collisions and in the absence of
thermal conduction.  In this case, the equilibrium timescale
is $t_{eq} = 5000 {\rm{yr}} E_{51}^{3/14} n_o^{-4/7}$, where
$E_{51}$ is the explosion energy in units of $10^{51}$~ergs
(Cox and Anderson 1982).  For Remnants A and B,
this formula gives an equilibration timescale
of 60,000~yrs and for Remnant C, it gives only 10,000~yrs more.
Plasma instabilities and electron and ion thermal
conduction will bring the particles into equilibrium faster.
Compared with the lifetimes of the SNRs, and especially when
compared with the lifetimes
of their energy conserving phases, this timescale is short.
While in
effect, the disequilibrium can change the bremsstrahlung power 
and spectral shape, ionization and recombination rates, and line emission.
Some of these changes may have longterm ramifications.

One must access a spectral model in order to calculate the radiative 
cooling rates and spectral
features.  In this case, the
Raymond and Smith (1977, 1993) spectral model was used.  Applying
others, such as those of Landini and Monsignori Fossi (1990),
Mewe {\it{et al}} (1985), Mewe {\it{et al}} (1986), 
Kaastra (1992), or Masai (1984) could lead to some differences in
the detailed spectra, though the differences 
in predicted broad band count rates should be small.
Note that the Raymond and Smith model was primarily designed for
calculations of collisional ionizational equilibrium, low energy resolution
spectra.  Better non collisional ionizational equilibrium emission 
calculations may
require inner-shell ionization and other processes not currently
written into the Raymond and Smith code.  The spectral
code community is strenuously working to improve the modeling 
and update the atomic constants.
% Currently, there are significant efforts to refine the
% atomic codes.
% and the improvements could lead to additional differences
%in the cooling rates and spectra.
Furthermore, because the hydrodynamic and spectral codes do not include 
cosmic ray acceleration, the predicted spectra
do not include X-ray synchrotron emission like that 
observed toward a few very young remnants (Koyama {\it{et al.}}
1995, Allen {\it{et al.}} 1997).

\section{Selected Applications}

\subsection{Application to the nearby lower Galactic halo}

Significant numbers of X-ray photons are produced above the
HI layer of the Galactic Disk.  
Snowden, {\it{et al.}}'s 1998 analysis found the 
de-absorbed ROSAT $1/4$~keV surface brightness
of the gas above the HI layer (after subtracting the extragalactic
flux) to be
400~counts~sec$^{-1}$~arcmin$^{-2}$ for
the south and 1150~counts~sec$^{-1}$~arcmin$^{-2}$ for the
north.  The average northern surface brightness is much more that
of the south because the north contains anomalous regions such 
as Loop I.

The high latitude sky is also rich in \oxyfive, \nitfour, and
\carthree\ (Shelton 1998, Savage, {\it{et al.}} 1997).
Determining the source of the X-ray emission and high stage ions
challenges researchers;
many physical scenarios are plausible
and multiple mechanisms may be at work.  One type of
contributor is the population of isolated SNRs originating above
the Galactic disk.  This subsection reports on the X-ray contribution
from those SNRs, for the case in which the explosion energy is
$10^{51}$~ergs and the halo environment is
tepid ($\sim10^4$~K) with a nonthermal pressure (magnetic and
cosmic ray combined) of 7200~K cm$^{-3}$.  The supernova remnant
simulations use $n = 0.01$~cm$^{-3}$, but preliminary estimates
show that during their lifetimes, SNRs in somewhat denser media produce
similar numbers of soft X-ray photons.  Hence the 
lifetime integrated results for this simulation (Remnant C) may be
used to approximate those of SNRs at a
range of heights above the Galactic disk.

In order to estimate the contributed X-ray surface brightness,
the simulation results must be combined with the 
statistical distribution of isolated supernova progenitors residing
at least 160~pc above or below the Galactic disk
(to avoid overlap with the region
occupied by the Local Bubble and to be above most of the
Galactic Disk's HI layer).
The progenitor rates are not precisely known and so here I will take
two rates as upper and lower estimates.  
The first set of isolated progenitor rates is that of 
Ferri\`{e}re (1995).  The second is a combination of
McKee and Williams (1997) massive star progenitors with
Ferri\`ere's Galactocentric radial distribution and her Type
1a rates (See Paper III).  With these rates,  
a population of Remnant C - like SNRs produces an
average ROSAT $1/4$~keV surface brightness of 230 to 
390~counts~sec$^{-1}$~arcmin$^{-2}$.
%{\bf{(confirmed)}}.
Given the uncertainties, this compares well with Snowden {\it{et al.'s}}
1998 observationally based estimate of the Southern Galactic
halo's surface
brightness.
%(400~counts~sec$^{-1}$~arcmin$^{-2}$).  It is smaller than
%their estimate for the Northern halo, but their estimate
%includes contributions from anomalous regions such as Loop I.

The image produced by a population of Remnant C - like SNRs is
consistent with the observations of the Galaxy's southern halo.
%, though
%bright features due to additional sources could also fit into the picture.
The population of Remnant C - like SNRs would appear as
a couple of bright (1000's of counts~sec$^{-1}$~arcmin$^{-2}$), 
%{\bf{(confirmed)}}
limb brightened regions and dozens of dim 
(100's of counts~sec$^{-1}$~arcmin$^{-2}$),
%{\bf{(confirmed)}},
centrally filled regions scattered across the high latitude sky.
The remnants are sufficiently plentiful and longlived as to cover
roughly half of the high latitude sky (including overlap of remnants).  
This compares well with
the observations of the southern halo, 
which can be described as a mottled ``background'' overlayed with
scattered bright features.  Paper III (Shelton 1999) provides 
greater detail and additional
calculations (such as the SNRs'
area coverage, and hot gas volume filling). 

Paper I shows the a population of Remnant A-like SNRs
produces as much \oxyfive, \nitfour, and \carthree\ as is observed
via absorption measurements toward stars
within the first kiloparsec of the plane.  This
paper shows that Remnant B and C-like SNRs produce similar or
larger quantities of these ions, thus extending the conclusions of Paper I 
to include these types of remnants.  
%Remnants B and C's high stage ions are also more emissive than those
%in Remnant A.  
An idea which escaped Paper I is the notion of buoyancy.  The
hot diffuse gas in the SNRs should rise.  As it does, it moves to
more tenuous and less pressurized surroundings.  Thus, the
remnant bubbles will further expand and their gas densities decrease.
With lower gas densities, the cooling and recombinations will 
slow.  The remnants will live longer, causing the time
integrals of the numbers of
\oxyfive, \nitfour, and \carthree\ per SNR to increase.  
The \carthree\ will be
especially effected.  Not only should this phenomena increase the
estimated number of high stage ions 
produced by the population of high latitude SNRs, 
but it should also increase the theoretical estimates of
the scale heights and ratio of \carthree\ to \oxyfive\ atoms.

\subsection{Application to the Local Hot Bubble}

%{\bf{(Mention that this is similar to the Breitschwerdt and Schmutzler
%model in having once heated and still highly ionized gas, but doesn't
%require the B and S expansion history.  Also, there is a Breitschwerdt
%et al 1996 article in Space Science Reviews, vol 78, issue 1/2, p. 183-198
%that I should look at.)}}

The Local Hot Bubble (LHB), also called the Local Bubble,
is a large (diameter $\sim 50$ to 100~pc), 
diffuse ($n \sim 0.05$ cm$^{-3}$), presumably hot ($T_c = 10^6$~K), 
X-ray emissive region containing the Sun (Snowden, {\it{et al}} 1990,
Warwick, {\it{et al.}} 1993, Cox and Reynolds 1987).  
The LHB is situated within a cavity called the
Local Cavity.  In some directions
the Local Cavity extends far beyond the X-ray emitting region
(Welsh {\it{et al.}} 1994, Snowden {\it{et al.}} 1998).
The Local Hot Bubble also contains several parsec-scale 
clouds and complexes of clouds,
including the Local Cloud complex which, as expected from its name,
surrounds the sun (Lallement 1998).
Although its existance has long been suspected, 
understanding the Local Hot Bubble's 
origin has proven to be a very challenging enterprise\footnote{The reader
will find a fascinating set of readings on this topic in ``The Local
Bubble and Beyond'', 1998, edited by Breitschwerdt, Freyberg, and Tr\"umper}.

The observations impose a host of difficult constraints on
LHB modeling.  
Snowden {\it{et al}} (1998) determined that the 
%of the ROSAT data partitions the observed $1/4$~keV flux between local
%and distant components.  The 
local region's $1/4$~keV surface brightness is
$\sim250$ to $\sim820 \times 10^{-6}$ counts s$^{-1}$ arcmin$^{-2}$
and the R1$/$R2 color temperature is $\sim10^6$~K.  
%Less emission is seen with the
The interior of the Local Bubble appears to produce little 
$3/4$~keV emission (Snowden, McCammon, and Verter 1993) and no evidence
has been found for an X-ray bright edge in either the
$1/4$ or $3/4$~keV band.
%ROSAT Wide Field Camera {\bf{(fill in photon energy range)}} than 
%is expected from the Be and B band data (not published).  
Furthermore, high spectral resolution observations by the 
Diffuse X-ray Spectrometer (Sanders {\it{et al.}} 1998)
force constraints on any detailed spectral model.
%150 to 284~eV spectra 

%At lower ionization levels, the Local Bubble is rich in \oxyfive.  
The average \oxyfive\ column density on
a sightline from the earth though the Local Cloud complex
and the Local Hot Bubble,
is $\sim1.6 \times 10^{13}$
\oxyfive\ cm$^{-2}$, with a velocity centroid near 0~km~s$^{-1}$
(Shelton \& Cox 1994). 
In order to explain the high column density of approximately 
stationary \oxyfive, either
the edge of the hot, highly ionized region is 
nearly stationary, most of the \oxyfive\ is located within the bubble's 
stationary interior, 
or the slow-moving Local Cloud is producing the observed \oxyfive\
and either 
the Local Hot Bubble's \oxyfive\ component
is moving faster than {\it{Copernicus}}'s velocity range 
($\sim \pm 100$~km~s$^{-1}$) or is negligibly small.
Simultaneously satisfying all of constraints is difficult.

The Remnant C simulations may be pointing the way to a promising
portion of parameter space. 
When Remnant C is several million years old, it
%shows that a single ancient SNR
%evolving in a diffuse, tepid, and high 
%nonthermal pressure environment can create 
has an X-ray emitting bubble of about the appropriate
size, emits copious $1/4$~keV X-rays, is not limb brightened in X-rays,
is fairly dim in $3/4$~keV X-rays, and contains plentiful quantities of 
stationary \oxyfive\ (the \oxyfive\ column density for a sightline looking
out from the center is $\sim 2.7 \times 10^{13}$ cm$^{-2}$).  
Like in the Breitschwerdt and Schmutzler (1994) model, 
the gas has cooled from a much
hotter temperature and contains overionized atoms.

Is the proposed physical scenario reasonable 
-- is the ambient medium relatively diffuse
and nonthermal pressure dominated?  Yes.  The Local Bubble is situated
inside the Local Cavity, a low density region.  If the Local Cavity
predates the Local Bubble, then the Local Bubble has been evolving
in a rarefied medium.  Furthermore,
if the Local Cavity gas is tepid and the total pressure is typical of the total
pressure in the Galactic midplane, 
then the pressure would be largely nonthermal.  
Thus, the proposed nature of the environment is reasonable.  
%{\bf{(left off here.  Added idea is that the snr is
%old and that agrees with the fact that there aren't any more
%O and B stars here.)}}
Some adjustments and further spectral testing of the model are needed, 
of course.  For example,
the local region is expected to
have a higher total pressure than Remnant C and so the author is
currently working on higher pressure models.  
%{\bf{(confirmed)}}

%If the Local Bubble is a SNR similar to 
%simulated Remnants A, B, or C during their pre-radiative phases, 
%then the X-ray emission is bright, but the \oxyfive\ has a large
%bulk velocity.   If the Local Bubble is a SNR similar to simulated
%Remnants A after the onset of shell formation, then
%the \oxyfive\ has a more reasonable bulk velocity, but the X-ray
%surface brightnesses are low.  
%
%However, by comparing the
%long term evolution of Remnant B (ambient $B_{eff} = 5 \mu$G) with
%that of Remnant A (ambient $B_{eff} = 2.5 \mu$G), it becomes clear that
%doubling the magnetic field has increased the X-ray luminosity
%in the post radiative phase by nearly an order of magnitude.
%At $3 \times 10^6$~yrs, Remnant B emits 
%$8.6 \times 10^{45}$ counts arcmin$^{2}$ s$^{-1}$ in the ROSAT $1/4$~keV 
%band and has a radius of 100~pc.  If seen from its center, it would be
%responsible for
%a $1/4$~keV countrate of $48 \times 10^{-6}$ 
%counts arcmin$^{-2}$ s$^{-1}$.
%This value is lower than the observationally determined rate, but
%might be additionally increased by increasing the explosion energy
%from the rather modest $0.5 \times 10^{51}$~ergs to 
%$1.0 \times 10^{51}$~ergs and increasing the
%nonthermal ambient pressure to a value more like that expected for
%the Galactic Disk.  Thus, while Remnants A and B do 
%C may not provide a
%one-to-one matches with the Local Bubble, they might be pointing
%the way toward a more promising region of parameter space.

\section{Summary}

It is difficult to do justice while condensing the life histories of the
three SNRs into a mere paragraph or two.   With that said, here is an attempt.
The gross features of their life histories are similar.
During their energy conserving phases, each remnant is hot, highly
pressured and rapidly expanding.  The relatively
dense gas swept up by the shockfront contributes nearly all of the
X-ray photons, causing the remnants to appear edge brightened.
The remnants are also very luminous, with surface brightnesses of
thousands of $1/4$~keV counts s$^{-1}$ arcmin$^{-2}$.
The ionization timescales significantly lag the dynamic timescales,
initially 
causing the gas to be drasticly underionized.  
%Thus, the X-ray spectra 
%exhibit emission lines from intermediate level ions.

Between 100,000 and 250,000 years, the remnants enter their radiative
phases.  The shockfront of each remnant slows 
to the extent that 
it is no longer able to dramaticly heat the gas it encounters and 
a cool shell develops between the shockfront and the
hot bubble.  
%For a while the hot bubbles continue to expand, but the
%pressure of the surrounding gas eventually stops the expansions.
Even without a very hot outer edge,
the hot bubbles continue to emit X-rays, 
but with lesser luminosities and without the strong edge brightening
of the young remnants.  
Remnants B and C are an order of magnitude brighter than Remnant A.
Compared with Remnant A, their larger 
ambient nonthermal pressures better compress the 
SNR bubbles, elevating the bubbles' temperatures and densities, and hence
X-ray luminosities.
Remnants B and C
have surface brightnesses of hundreds of 
$1/4$~keV counts s$^{-1}$ arcmin$^{-2}$ 
during their old evolutionary phases.
(Remnants B and C produce more $1/4$~keV photons during their old evolutionary
phases than during their youth.)
The recombination timescale
lags the cooling timescale, causing the atoms to become overionized
around the time that the cool shell forms and to remain overionized 
until the hot bubbles disappear, some 12 million or more years later.
The simulation parameters, lifetimes, sizes, $1/4$~keV, $3/4$~keV output, 
and the numbers of \oxyfive,
\nitfour, and \carthree\ atoms are compiled into Table 1.

	An effort was made to understand how such remnants would appear if
observed with modern or future facilities and analyzed with common techniques
which assume that the ions are in collisional equilibrium with respect to
the gas temperature. 
The ``color temperature'' was found from the ratios of the ROSAT PSPC
R1, R2, and R4 band countrates, 
the electron density and thermal pressures were calculated from
the color temperatures and surface brightnesses, and the O VII and
O VIII ratios were compared with those of collisional equilibrium plasma.
Remnant A was used as an example.
%The color temperature (calculated from ratios of various ROSAT PSPC
%bands) slowly varies and rarely matches the kinetic
%or ionization temperature of the emitting gas.  
When seen through the
lens of these techniques, the young remnant appears,
denser, and less pressurized than it is
and the old remnant appears
hotter, more rarefied, and less pressurized than it is.  The O VII and
O VIII diagnostics accurately identify the plasma in the young remnant
as being underionized and the
plasma and the old remnant as containing overionized, recombining gas.

	The results were 
%parameterized for comparison with observations of external galaxies, 
combined with the Galaxy's progenitor statistics
in order to compare with the observed $1/4$~keV surface
brightness of the Galaxy's southern halo, and analyzed for clues as to
the origin of the Local Hot Bubble.  In the case of the Galactic halo,
the number of type O and B runaway stars and type Ia progenitors
which explode a few hundred 
parsecs from the disk is remarkably large, as is the 
number of soft X-ray photons emitted during Remnant C's lifetime.  Combining
these values shows that a population of Remnant C-like remnants could 
explain roughly 200 to 400 counts s$^{-1}$ arcmin$^{-2}$ of the 
observed 400 counts s$^{-1}$ arcmin$^{-2}$ in the
ROSAT PSPC $1/4$~keV band.  These SNRs can also
explain the observed spatial emission pattern which consists of
a few bright regions, a dim
mottled background, and about half of the high latitude southern
sky having nothing but the
local and extragalactic fluxes.  In the case of the Local Hot Bubble,
a one million year old 
Remnant C comes within sight of explaining the Local Hot Bubble's
size, $1/4$~keV surface brightness, lack of limb brightening, 
and column density of $\sim0$~km~s$^{-1}$ bulk velocity \oxyfive\ ions.

%{\bf{(confirmed ed)}}.
%varies slowly is a poor indicator of either the ionization
%or the kinetic temperature.  During the snowplow phase, the color
%temperature is drastically lower than the kinetic temperature and later in
%the evolution, the color temperature is consistently higher than the
%kinetic temperature.
%{\bf{(consider removing: For comparison with future observations,
%plasma diagnostics such as the O VII and O VIII line strengths and
%the fraction of energy emitted in line radiation versus
%continuum emission are analyzed.
%For the first hundred thousand years, the O VII ``triplet'' and the ratio of
%the O VII to O VIII flux follow the expected trajectory for ionizing
%plasma.  Later, the atoms are recombining, which may
%be difficult to determine from moderate energy resolution spectra unless
%the recombination lines are also considered.  In general, the line to
%continuum ratios are found to be very poor diagnostics and should not
%be used to find the gas phase abundances.)}}
%The fraction of the explosion's 
%kinetic energy which is eventually 
%radiated away in the form of the soft X-ray photons is found to be 
%about $2\%$.

\vspace{1cm}

\noindent{\bf{Acknowledgements}}

While at the University of Wisconsin, Department of Physics, the
author received invaluable assistance from Don Cox and thanks
him for sharing his hydrocode expertise, astrophysics intuition, and
good cheer.  While at the NASA/Goddard Space Flight Center, Laboratory
for High Energy Astrophysics, the author received irreplaceable
assistance from Rob Petre, Steve Snowden, and Kip Kuntz, and thanks
them for sharing their X-ray analysis expertise, bibliographic
memories, and moral support.
This work was supported under NASA grant NAG5-3155 and by an award from
the Wisconsin Space Grant Consortium while the author was at the
University of Wisconsin -- Madison, and by a grant from the
National Research Council while the author was at the 
NASA/Goddard Space Flight Center, Laboratory
for High Energy Astrophysics.

\clearpage

%are presented in Table~\ref{table1}.
%\vspace{.2cm}

\begin{table}
\begin{center}
\begin{tabular}{|c|c|c|c|}
\hline \hline       & Remnant A & Remnant B  & Remnant C \\ \hline \hline 
\ ambient $n$       & 0.01 cm$^{-3}$ & 0.01 cm$^{-3}$ & 0.01 cm$^{-3}$ \\ \hline 
\ ambient $T$       & $1.0 \times 10^4$~K & $1.0 \times 10^4$~K & $1.0 \times 10^4$~K \\ \hline 
\ ambient $B$       & 2.5 $\mu$G & 5 $\mu$G & 5 $\mu$G \\ \hline 
\ explosion energy  & $0.5 \times 10^{51}$ ergs & $0.5 \times 10^{51}$ ergs & $1.0 \times 10^{51}$ ergs \\ \hline 
\ lifetime          & 16 million yrs & 12 million yrs & 14 million yrs \\ \hline 
\ maximum radius    & 140 pc & 110 pc & 140 pc \\ \hline 
\ $1/4$~keV output  & $5.8 \times 10^{59}$ counts cm$^{2}$ & $1.4 \times 10^{60}$ counts cm$^{2}$ & $3.5 \times 10^{60}$ counts cm$^{2}$ \\ \hline 
\ $3/4$~keV output  & $1.2 \times 10^{58}$ counts cm$^{2}$ & $1.1 \times 10^{58}$ counts cm$^{2}$ & $2.8 \times 10^{58}$ counts cm$^{2}$ \\ \hline 
\ \oxyfive          & $7.8 \times 10^{69}$ \oxyfive\ sec & $6.0 \times 10^{69}$ \oxyfive\ sec & $1.2 \times 10^{70}$ \oxyfive\ sec \\ \hline 
\ \nitfour          & $7.2 \times 10^{68}$ \nitfour\ sec & $4.6 \times 10^{68}$ \nitfour\ sec & $8.9 \times 10^{68}$ \nitfour\ sec \\ \hline 
\ \carthree         & $1.6 \times 10^{69}$ \carthree\ sec & $1.1 \times 10^{69}$ \carthree\ sec & $2.4 \times 10^{69}$ \carthree\ sec \\ \hline \hline
\end{tabular}
\end{center}
\caption[]
{Simulation parameters, lifetimes, maximum radius of the hot bubbles, 
and the time integrated numbers of soft X-ray counts and high stage ions
for Remnants A, B, and C.}
\label{table1}
\end{table}

%  put this in the halo paper
%\begin{figure}           
%%  \figcaption[map of the 1/4 keV X-ray emission from the distant
%%component]
%  \caption[map of the 1/4 keV X-ray emission from the distant
%component]
%{Maps of the distant $1/4$~keV X-ray surface brightness.  ({\it{a}})
%Northern pole:  $l = 0^{\rm{o}}$ is at the bottom, 
%$l = 90^{\rm{o}}$ as to the left, 
%$l = 180^{\rm{o}}$ as at the top, 
%$l = 270^{\rm{o}}$ as to the right.  Latitude lines are drawn for
%every $15{\rm{o}}$.
%({\it{b}}) Southern pole:  $l = 0^{\rm{o}}$ is at the top, 
%$l = 90^{\rm{o}}$ as to the left, 
%$l = 180^{\rm{o}}$ as at the bottom, 
%$l = 270^{\rm{o}}$ as to the right.  Latitude lines are drawn for
%every $15{\rm{o}}$. This figure is from Snowden, {\it{et al.}} 1998.}  
%\label{fig:Snowden}
%\end{figure}

\begin{figure}           
%  \figcaption[Temperature versus Supernova Remnant Radius]
  \caption[Temperature versus Supernova Remnant Radius]
{Kinetic temperature of the gas versus radius from the center of Remnant A.
Figure {\it{a}} shows the first six epochs and demonstrates
that the effects of cooling become apparent
by $2.5 \times 10^5$ yrs.  
Figure {\it{b}} shows the subsequent six epochs.  In this figure
it can be seen that the hot bubble reaches its maximum size 
around $2 \times 10^6$ yrs and gradually shrinks afterwards.  
Figures {\it{c}} and {\it{d}} 
demonstrate the remnant's gradual cooling and 
collapse during the subsequent $1.2 \times 10^7$ yrs and its 
disappearance between $1.6$ and $1.7 \times 10^7$ yrs.}  
\label{temperature}
\end{figure}

\begin{figure}           %[tb]
%  \figcaption[Volume Density versus Supernova Remnant Radius]
  \caption[Volume Density versus Supernova Remnant Radius]
{Remnant A:  Volume density versus radius for the same 24 epochs
as in Figure~\ref{temperature}.}
\label{density}
\end{figure}

%\pagebreak

\begin{figure}           %[tb]
%  \figcaption[Thermal and Total Pressure versus Supernova Remnant Radius]
  \caption[Thermal and Total Pressure versus Supernova Remnant Radius]
{Remnant A:  The total and thermal pressure of the gas versus radius.  
Figure {\it{a}} is composed of
a plot of the total pressures for the first six epochs placed above
a plot of the thermal pressures for these epochs.  
Figures {\it{b}}, {\it{c}}, and {\it{d}} 
plot the remaining epochs in the same fashion.
During the earliest epochs, the thermal pressure contributes nearly all of
the total pressure, but by $1 \times 10^6$ 
the pressure in the shell is almost entirely non-thermal.}
\label{pressure}
\end{figure}

%\clearpage

\begin{figure}           %[tb]
%  \centerline{\vbox{\epsfbox{~/data2/halo/figures/halo_19profiles.ps}}}
  \caption{Remnant A's ionization and kinetic temperatures versus radius.  
The ionization temperature is 
marked with a solid line.  It was found from the ratio of 
Si$^{+8}$/Si$^{+9}$ ions, which
is a good indicator for $T_i$'s between 
$\sim6 \times 10^5$~K and a few times $10^6$~K.  The 
O$^{+6}$/O$^{+7}$ and O$^{+5}$/O$^{+6}$ ratios (dashed line) were used to
extend the ionization temperature curves down to $3 \times 10^5$~K.
%Where this is done, a dashed line is used.
%The O$^{+5}$/O$^{+6}$ ratio 
%is not a good temperature indicator for 
%$T_i \stackrel{<}{\sim} \sim3 \times 10^5$~K plasma and so the 
%curves are terminated at this point.
The kinetic temperature is marked with a dotted line.  
The upper panel (a) displays the curves for
(A) $10^4$, 
(B) $2.5 \times 10^4$, 
(C) $5 \times 10^4$, 
(D) $10^5$, and
(E) $2.5 \times 10^5$~yrs, 
while the lower panel (b) displays the curves for 
(F) $10^6$, 
(G) $5 \times 10^6$, 
(H) $10^7$, and 
(I) $1.5 \times 10^7$~yrs.}
\label{fig:iont}
\end{figure}

%\pagebreak

%\begin{figure}           %[tb]
%%  \centerline{\vbox{\epsfbox{~/data2/halo/figures/halo_19profiles.ps}}}
%  \caption{The total luminosity 
%(per eV) between 50 and 1150~eV from Remnant A
%as a function of photon energy.  Plot (a) includes the
%spectrum for 10,000 yrs shifted by $2 \times 10^{35}$ ergs s$^{-1}$ eV$^{-1}$, 
%spectrum for 25,000 yrs shifted by $1.5 \times 10^{35}$ ergs s$^{-1}$ 
%eV$^{-1}$, 
%spectrum for 50,000 yrs shifted by $1 \times 10^{35}$ ergs s$^{-1}$ eV$^{-1}$, 
%spectrum for 100,000 yrs shifted by $0.5 \times 10^{35}$ ergs s$^{-1}$ 
%eV$^{-1}$, and the
%spectrum for 250,000 yrs.
%Plot (b) repeats the spectrum for 250,000 yrs, but shifted by 
%$10 \times 10^{31}$ ergs s$^{-1}$ and includes the 
%spectrum for 1,000,000 yrs shifted by $7.5 \times 10^{31}$ ergs s$^{-1}$ eV$^{-1}$, 
%spectrum for 5,000,000 yrs shifted by $5 \times 10^{31}$ ergs s$^{-1}$ eV$^{-1}$, 
%spectrum for 10,000,000 yrs shifted by $2.5 \times 10^{31}$ ergs s$^{-1}$ eV$^{-1}$, and the 
%spectrum for 15,000,000 yrs.}
%\label{fig:spectralinear}
%\end{figure}

%\clearpage
%\pagebreak

\begin{figure}           %[tb]
%  \centerline{\vbox{\epsfbox{~/data2/halo/figures/halo_19profiles.ps}}}
  \caption{The logarithm of the total luminosity 
(per eV) between 50 and 1150~eV from remnant A 
as a function of photon energy.  Several epochs are plotted, with all
but the first scaled to aid legibility.
From top to bottom, the spectra are for: 
$10^4$~yrs (unscaled), 
$2.5 \times 10^4$~yrs (scaled by $10^{-4}$), 
$5 \times 10^4$~yrs (scaled by $10^{-8}$), 
$10^5$~yrs (scaled by $10^{-12}$), 
$2.5 \times 10^5$~yrs (scaled by $10^{-16}$), 
$10^6$~yrs (scaled by $10^{-20}$), 
$5 \times 10^6$~yrs (scaled by $10^{-24}$), 
 $10^7$~yrs (scaled by $10^{-28}$), 
and 
$1.5 \times 10^7$~yrs (scaled by $10^{-32}$).}
\label{fig:spectralog}
\end{figure}

%\begin{figure}
%  \caption{Remnant A's 
%time integrated radiated energy as a function of photon energy.
%The solid line tracks the emission from the remnant in a 
%$n_o = 0.01$ cm$^{-3}$,
%$B_{eff} = 2.5\mu$G ambient medium (Remnant A) and the dotted
%line tracks the emission from the remnant in a $n_o = 0.01$ cm$^{-3}$,
%$B_{eff} = 5.0\mu$G ambient medium (Remnant B).  
%{\bf{(Add Remnant C.)}} The radiated
%energies were
%calculated using 10eV bins in photon energy and later normalized
%to determine the energy emitted per 1eV.  This has slightly dulled the
%emission lines and recombination edges.}
%\label{fig:cumulative}
%\end{figure}

%\clearpage
%\pagebreak

\begin{figure}           %[tb]
%  \figcaption[]
  \caption[]
{Band-response functions for the ROSAT PSPC R1, R2, R4, and R5 bands,
courtesy of Steve Snowden.}
% This is the newer figure from Steve, it is in figures/bandresponse.ps.
\label{bandfunctions}
\end{figure}

%\pagebreak

\begin{figure}
  \caption{Luminosity of Remnant A in units of $10^{40}$
ROSAT PSPC $1/4$ and $3/4$~keV
band counts cm$^2$ s$^{-1}$ as a function of age.
The luminosity peaks first while the remnant is in its energy
conserving phase.  It peaks a second time during the
remnant's collapse.}  
\label{fig:totalxray}
\end{figure}

\begin{figure}
  \caption{Remnant A's 
ROSAT PSPC $1/4$~keV band countrate versus
impact parameter.  The flux is in units of $10^{-6}$ counts
s$^{-1}$ arcmin$^{-2}$.  While the remnant is in its
energy conserving phase, it is bright and has a strongly limb brightened
appearance.  
Afterward, it is dim and has a centrally filled appearance.}  
\label{fig:R1flux}
\end{figure}

%\clearpage
%\pagebreak

%\begin{figure}
%  \caption{Remnant A's
%ROSAT PSPC R2 band ($\sim140$ to 284 eV) countrate versus
%impact parameter.  The flux is in units of $10^{-6}$ counts
%s$^{-1}$ arcmin$^{-2}$.}  
%\label{fig:R2flux}
%\end{figure}

%\pagebreak

\begin{figure}
  \caption{Remnant A's ROSAT PSPC $3/4$~keV band 
%($\sim440$ to $\sim1010$ eV) 
countrate versus impact parameter.  The flux is in units of $10^{-6}$ counts
s$^{-1}$ arcmin$^{-2}$.}  
\label{fig:R4flux}
\end{figure}

%\pagebreak

%\begin{figure}           
%  \caption[]
%{Fractions of energy emitted by line and continuum processes:
%the solid line is the fraction of the total radiated energy which is
%emitted as line radiation.  The dashed
%line is the fraction of the total radiated energy which is
%emitted as continuum radiation.
%The fractions for the simulated
%remnant are plotted versus SNR age in 
%Figures {\it{a, b, c, d}}, and {\it{e}}, while 
%the fractions for an equilibrium plasma and are plotted
%versus plasma temperature
%in Figures {\it{f, g, h, i,}} and {\it{j}}.
%Figures {\it{a}} and {\it{f}} are for the 100 to 300~eV range,
%Figures {\it{b}} and {\it{g}} are for the 300 to 500~eV range,
%Figures {\it{c}} and {\it{h}} are for the 500 to 700~eV range,
%Figures {\it{d}} and {\it{i}} are for the 700 to 900~eV range,
%Figures {\it{e}} and {\it{j}} are for the 900 to 1100~eV range.}
%\label{ratio}
%\end{figure}
%
%\pagebreak

\begin{figure}           
%  \figcaption[Temperature versus Supernova Remnant Radius]
  \caption[Temperature versus Supernova Remnant Radius]
{Kinetic temperature of the gas versus radius for Remnant B.
Figure {\it{a}} shows the first set of six epochs,
%By $2.5 \times 10^5$ yrs, the effects of cooling have become apparent.  
figure {\it{b}} shows the subsequent set of six epochs, and Figure
{\it{c}} shows the third set.  By $1.3 \times 10^7$~yrs, 
the remnant has completely disappeared.}  
\label{temperature5uG}
\end{figure}

%\clearpage

\begin{figure}           %[tb]
%  \figcaption[Volume Density versus Supernova Remnant Radius]
  \caption[Volume Density versus Supernova Remnant Radius]
{Remnant B:  Volume density versus radius for the same 18 epochs
as in Figure~\ref{temperature5uG}.}
\label{density5uG}
\end{figure}

\begin{figure}           %[tb]
  \caption[Thermal and Total Pressure]
{Remnant B: The thermal and total pressure of the gas is plotted 
versus radius.}
\label{pressure5uG}
\end{figure}

%\begin{figure}           %[tb]
%%  \centerline{\vbox{\epsfbox{~/data2/halo/figures/halo_29profiles.ps}}}
%  \caption{The total luminosity 
%(per eV) between 50 and 1150~eV from Remnant B 
%as a function of photon energy.  Plot (a) includes the
%spectrum for 10,000 yrs shifted by $2 \times 10^{35}$ ergs s$^{-1}$ eV$^{-1}$, 
%spectrum for 25,000 yrs shifted by $1.5 \times 10^{35}$ ergs s$^{-1}$ 
%eV$^{-1}$, 
%spectrum for 50,000 yrs shifted by $1 \times 10^{35}$ ergs s$^{-1}$ eV$^{-1}$, 
%spectrum for 100,000 yrs shifted by $0.5 \times 10^{35}$ ergs s$^{-1}$ 
%eV$^{-1}$, and the
%spectrum for 250,000 yrs.
%Plot (b) repeats the spectrum for 250,000 yrs, but shifted by 
%$10.5 \times 10^{31}$ ergs s$^{-1}$ and includes the 
%spectrum for 1,000,000 yrs shifted by $7 \times 10^{31}$ ergs s$^{-1}$ eV$^{-1}$, the
%spectrum for 5,000,000 yrs shifted by $3.5 \times 10^{31}$ ergs s$^{-1}$ eV$^{-1}$, and the
%spectrum for 10,000,000 yrs.}
%\label{fig:spectralinearB}
%\end{figure}

%\clearpage

\begin{figure}           %[tb]
%  \centerline{\vbox{\epsfbox{~/data2/halo/figures/halo_29profiles.ps}}}
  \caption{The logarithm of the total luminosity 
(per eV) between 50 and 1150~eV from Remnant B
as a function of photon energy.  Several epochs are plotted, with all
but the first scaled to aid legibility.
From top to bottom, the spectra are: 
$10^4$~yrs (unscaled), 
$2.5 \times 10^4$~yrs (scaled by $10^{-4}$), 
$5 \times 10^4$~yrs (scaled by $10^{-8}$), 
$10^5$~yrs (scaled by $10^{-12}$), 
$2.5 \times 10^5$~yrs (scaled by $10^{-16}$), 
$10^6$~yrs (scaled by $10^{-20}$), 
$5 \times 10^6$~yrs (scaled by $10^{-24}$), 
and $10^7$~yrs (scaled by $10^{-28}$).
%and (I) $1.5 \times 10^7$~yrs (scaled by $10^{-32}$).  
%The spectra was calculated for 2~eV energy bins, then normalized.
}
\label{fig:spectralogB}
\end{figure}

%\clearpage

\begin{figure}
  \caption{Luminosity of Remnant B in units of $10^{40}$
ROSAT PSPC $1/4$ and $3/4$~keV
band counts cm$^2$ s$^{-1}$ as a function of age.  Remnant B
is similar to Remnant A for the first $10^5$~yrs, but is much brighter
at later times.}
\label{luminosityB}
\end{figure}

%\begin{figure}
%  \caption{Fraction of the total radiated energy which has been
%emitted with photon energies which are higher than photon
%energy plotted on the x axis.  The solid line is for Remnant A,
%while the dotted line is for Remnant B.}
%\label{fig:running}
%\end{figure}

\begin{figure}
  \caption{Remnant B's ROSAT PSPC $1/4$~keV band 
countrate versus impact parameter.  The flux is in units of $10^{-6}$ counts
s$^{-1}$ arcmin$^{-2}$.  Like the other remnants, Remnant B appears
edge brightened in its youth then evolves to appear centrally filled.}  
\label{fig:halo29R1flux}
\end{figure}

%\begin{figure}
%  \caption{Remnant B's ROSAT PSPC R2 band count 
%rate versus impact parameter.  The flux is in units of $10^{-6}$ counts
%s$^{-1}$ arcmin$^{-2}$.}  
%\label{fig:halo29R2flux}
%\end{figure}

%\clearpage

\begin{figure}
  \caption{Remnant B's ROSAT PSPC $3/4$~keV band 
countrate versus impact parameter.  The flux is in units of $10^{-6}$ counts
s$^{-1}$ arcmin$^{-2}$.}  
\label{fig:halo29R4flux}
\end{figure}

\begin{figure}           
%  \figcaption[Temperature versus Supernova Remnant Radius]
%[Temperature versus Supernova Remnant Radius]
\caption{Kinetic temperature of the gas versus radius from the center of the
remnant for Remnant C.
Figure {\it{a}} shows the first set of six epochs,
%By $2.5 \times 10^5$ yrs, the effects of cooling have become apparent.  
figure {\it{b}} shows the subsequent set of six epochs, Figure
{\it{c}} shows the third set, and Figure {\it{d}} shows the
1.3, and $1.4 \times 10^7$~yr epochs.}  
\label{fig:temperatureC}
\end{figure}

\clearpage

\begin{figure}           %[tb]
%  \figcaption[Volume Density versus Supernova Remnant Radius]
  \caption[Volume Density versus Supernova Remnant Radius]
{Remnant C:  Density of atoms versus radius.}
\label{fig:densityC}
\end{figure}

\begin{figure}           %[tb]
  \caption[Thermal and Total Pressure]
{Remnant C:  The thermal and total pressure of the gas is plotted 
versus radius for Remnnant.}
\label{fig:pressureC}
\end{figure}

%\clearpage

%\begin{figure}           %[tb]
%%  \centerline{\vbox{\epsfbox{~/data2/halo/figures/halo_29profiles.ps}}}
%  \caption{The total luminosity 
%(per eV) between 50 and 1150~eV from Remnant C 
%as a function of photon energy.  Plot (a) includes the
%spectrum for 10,000 yrs shifted by $3.6 \times 10^{35}$ ergs s$^{-1}$ eV$^{-1}$, 
%spectrum for 25,000 yrs shifted by $2.7 \times 10^{35}$ ergs s$^{-1}$ 
%eV$^{-1}$, 
%spectrum for 50,000 yrs shifted by $1.8 \times 10^{35}$ ergs s$^{-1}$ eV$^{-1}$, 
%spectrum for 100,000 yrs shifted by $0.9 \times 10^{35}$ ergs s$^{-1}$ 
%eV$^{-1}$, and the
%spectrum for 250,000 yrs.
%Plot (b) repeats the spectrum for 250,000 yrs, but shifted by 
%$12 \times 10^{32}$ ergs s$^{-1}$ and includes the 
%spectrum for 1,000,000 yrs shifted by $9 \times 10^{32}$ ergs s$^{-1}$ eV$^{-1}$, 
%spectrum for 5,000,000 yrs shifted by $6 \times 10^{32}$ ergs s$^{-1}$ eV$^{-1}$, 
%spectrum for 10,000,000 yrs shifted by $3 \times 10^{32}$ ergs s$^{-1}$ eV$^{-1}$, and the 
%spectrum for 15,000,000 yrs.}
%\label{fig:spectralinearC}
%\end{figure}

\begin{figure}           %[tb]
%  \centerline{\vbox{\epsfbox{~/data2/halo/figures/halo_29profiles.ps}}}
  \caption{The logarithm of the total luminosity 
(per eV) between 50 and 1150~eV from Remnant C
as a function of photon energy.  Several epochs are plotted, with all
but the first scaled to aid legibility.
From top to bottom, the spectra are for: 
$10^4$~yrs (unscaled), 
$2.5 \times 10^4$~yrs (scaled by $10^{-4}$), 
$5 \times 10^4$~yrs (scaled by $10^{-8}$), 
$10^5$~yrs (scaled by $10^{-12}$), 
$2.5 \times 10^5$~yrs (scaled by $10^{-16}$), 
$10^6$~yrs (scaled by $10^{-20}$), 
$5 \times 10^6$~yrs (scaled by $10^{-24}$), 
$10^7$~yrs (scaled by $10^{-28}$), 
and 
$1.5 \times 10^7$~yrs (scaled by $10^{-32}$).}  
\label{fig:spectralogC}
\end{figure}

%\clearpage

\begin{figure}
  \caption{Luminosity of Remnant C in units of $10^{40}$
ROSAT PSPC $1/4$ and $3/4$~keV
band counts cm$^2$ s$^{-1}$ as a function of age. Remnant C's luminosity
follows the pattern of 
Remnant B but is about two and a half times as bright.}  
\label{fig:luminosityC}
\end{figure}

\begin{figure}
  \caption{Remnant C's ROSAT PSPC $1/4$~keV band countrate 
versus impact parameter.  The flux is in units of $10^{-6}$ counts
s$^{-1}$ arcmin$^{-2}$.  Remnant C's appearance is similar to but
brighter than that of Remnant B.}  
\label{fig:halo31R1flux}
\end{figure}

%\begin{figure}
%  \caption{Remnant C's ROSAT PSPC R2 band count 
%rate versus impact parameter.  The flux is in units of $10^{-6}$ counts
%s$^{-1}$ arcmin$^{-2}$.}  
%\label{fig:halo31R2flux}
%\end{figure}

%\clearpage

\begin{figure}
  \caption{Remnant C's ROSAT PSPC $3/4$~keV band 
countrate versus impact parameter.  The flux is in units of $10^{-6}$ counts
s$^{-1}$ arcmin$^{-2}$.}  
\label{fig:halo31R4flux}
\end{figure}

\begin{figure}
  \caption{The top panel (a) shows the color temperature versus impact
parameter for Remnant A at (A) $10^4$, (B) $2.5 \times 10^4$, 
(C) $5 \times 10^4$, (D) $10^5$, and (E) $2.5 \times 10^5$~yrs.  The bottom
panel (b) is for 
(F) $10^6$, (G) $5 \times 10^6$, (H) $10^7$, and (I) $1.5 \times 10^7$~yrs.
The color temperature was found from the ROSAT PSPC R1 and
R2 bands (solid line) and the R2 and R4 bands (dotted line).
Note that the R1/R2 ratio only gives a unique
color temperature for $T_{cR1R2}$ below $\sim10^6$~K, so  
plotted values of $10^6$~K represent lower limits.  The
plotted values of the R2/R4 color temperature of 
$2.5 \times 10^6$~K also formally represent lower limits, but a detailed
examination of both band ratios gives the true R2/R4 color
temperature, which for Remnant A rarely exceeds $2.5 \times 10^6$~K.
Because the abscissa is impact parameter rather than
radius, these plots cannot be {\it{formally}} compared with 
Figures~\ref{temperature}, \ref{fig:R1flux}, 
or \ref{fig:R4flux}.}
\label{fig:colort}
\end{figure}

%\clearpage

\begin{figure}
  \caption{Remnant A's luminosity as a function of time in
the (a) O VII 561~eV (forbidden), (b) O VII 569~eV (intercombination), 
(c) O VII 571~eV (satellite), 
(d) O VII 574~eV (resonance), and (e) O VIII 653~eV (Lyman $\alpha$)
lines.  The late-time luminosity in the 574 eV line is not viewable on
the plot.  Sample values are:
$1.5 \times 10^{29}$ ergs cm$^{-1}$ at $1 \times 10^{6}$ yrs, 
$1.2 \times 10^{28}$ ergs cm$^{-1}$ at $5 \times 10^{6}$ yrs, and 
$5 \times 10^{26}$ ergs cm$^{-1}$ at $10 \times 10^{6}$ yrs.}  
\label{fig:ovii}
\end{figure}

\begin{figure}
  \caption{Remnant A:  ratio of the 
O VII forbidden line luminosity to the sum of the  
O VII resonance and intercombination line luminosities plotted versus the
ratio of the O~VIII~Lyman $\alpha$ 
line luminosity to the sum of the O VII resonance and
intercombination line luminosities (solid line).  
For comparison, the ratios for plasma in
collisional ionizational equilibrium are also plotted (dotted line).}
\label{fig:oviioviii}
\end{figure}

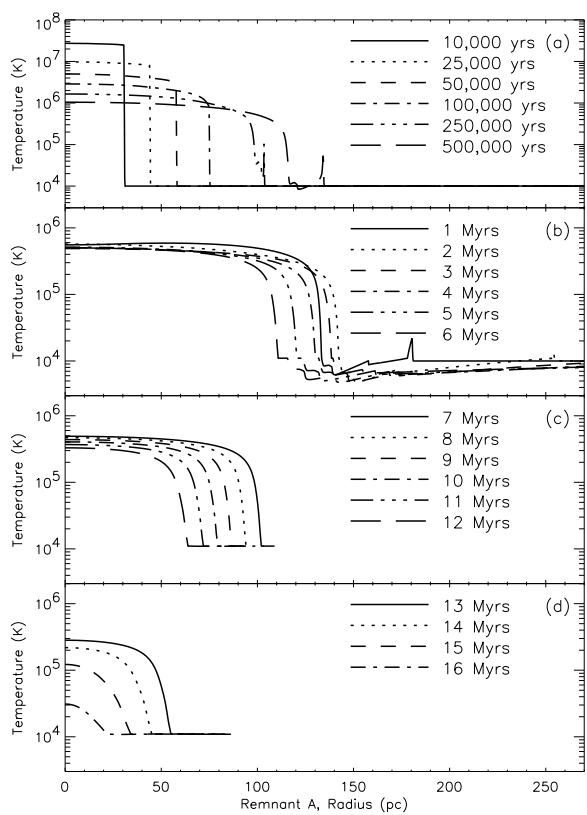
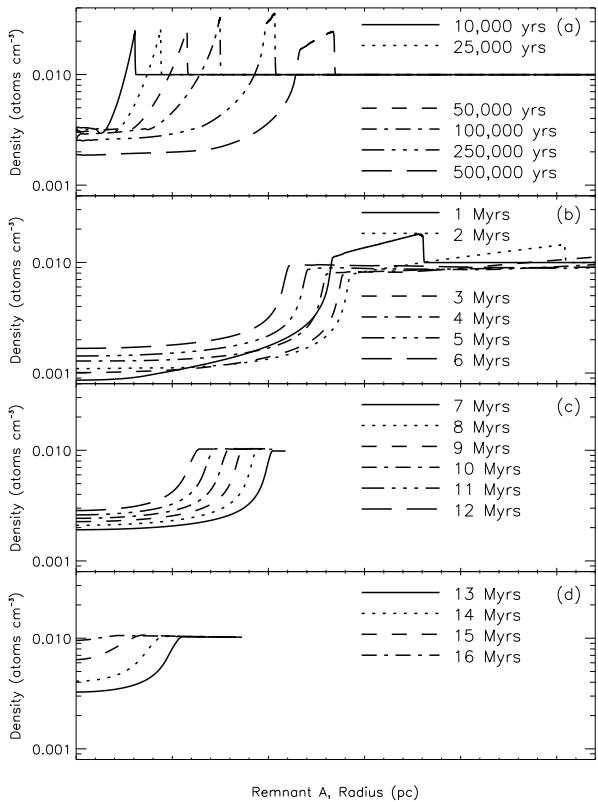
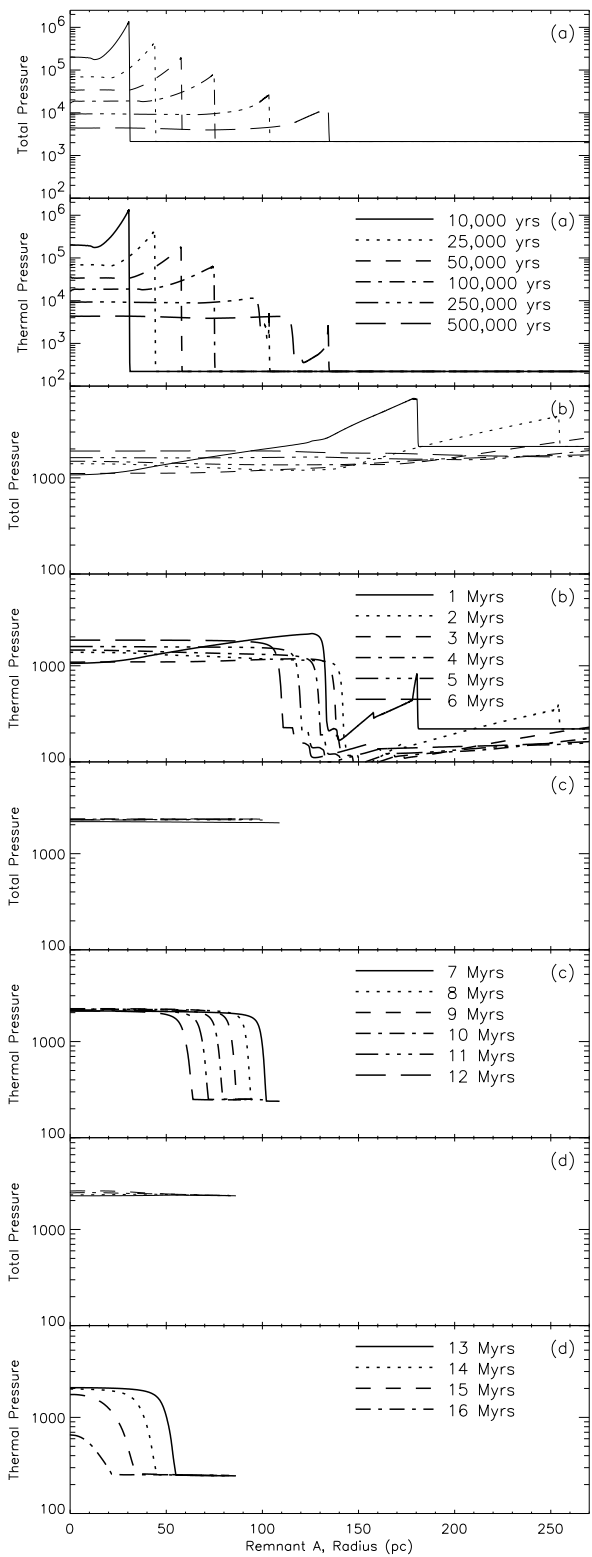
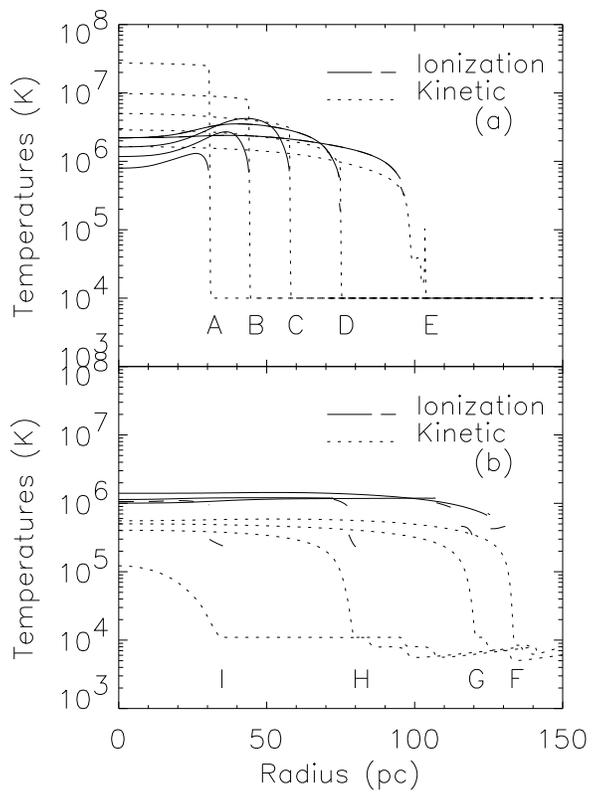
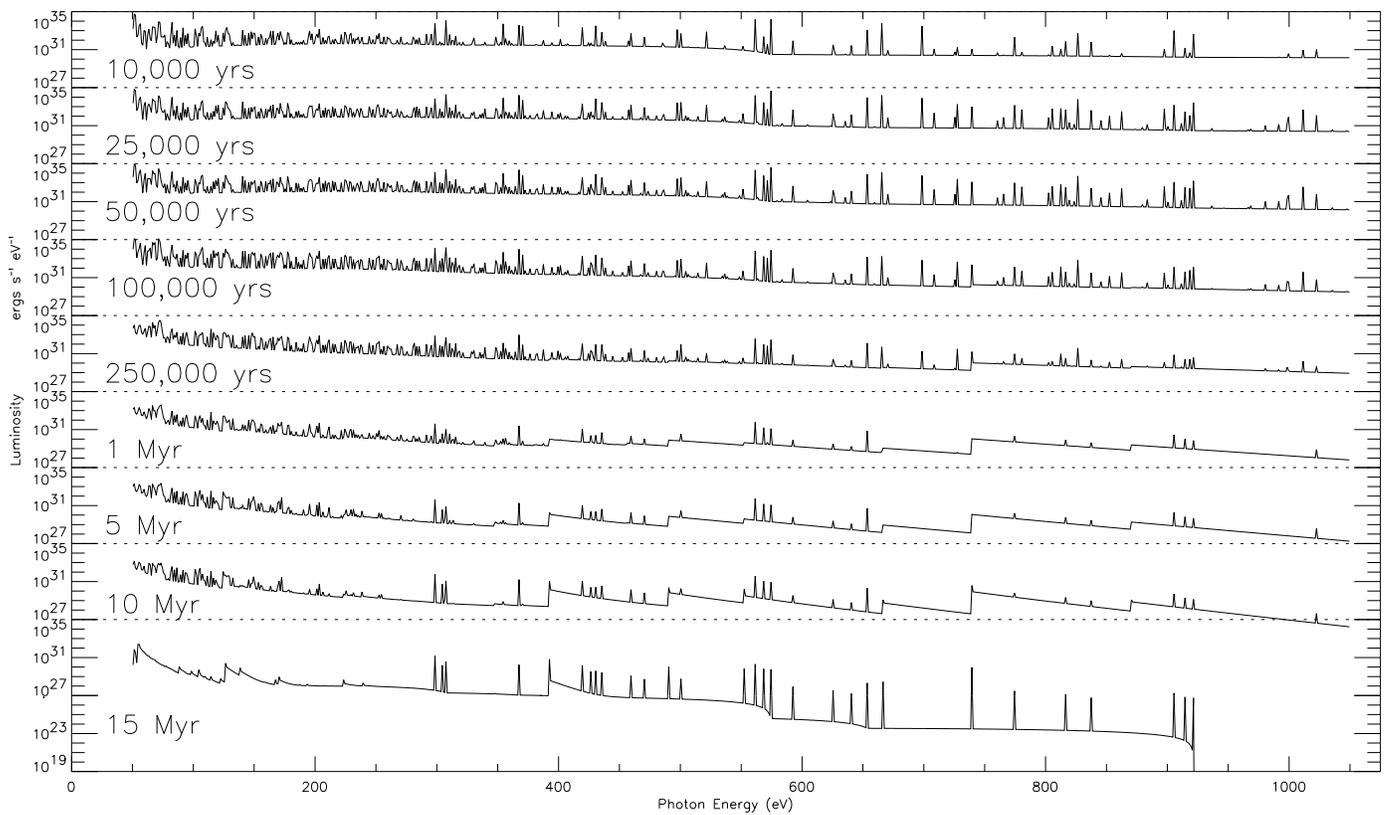
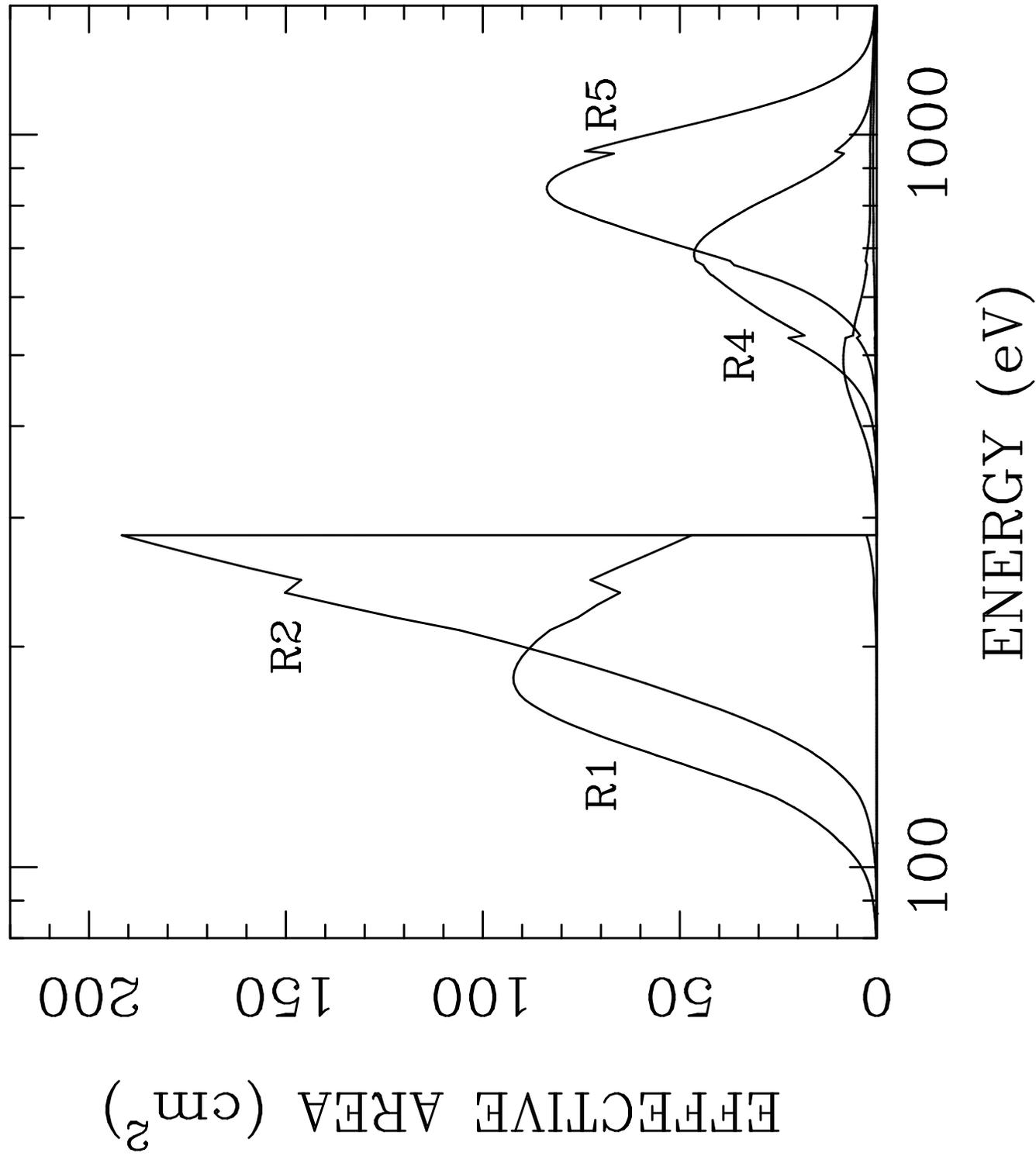
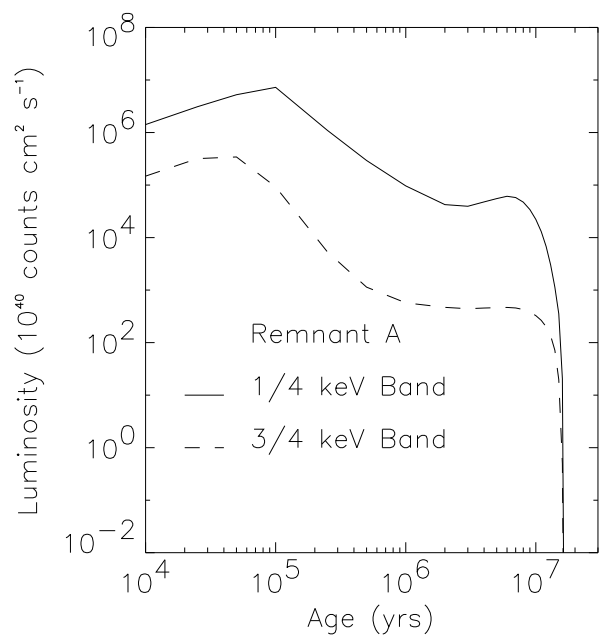
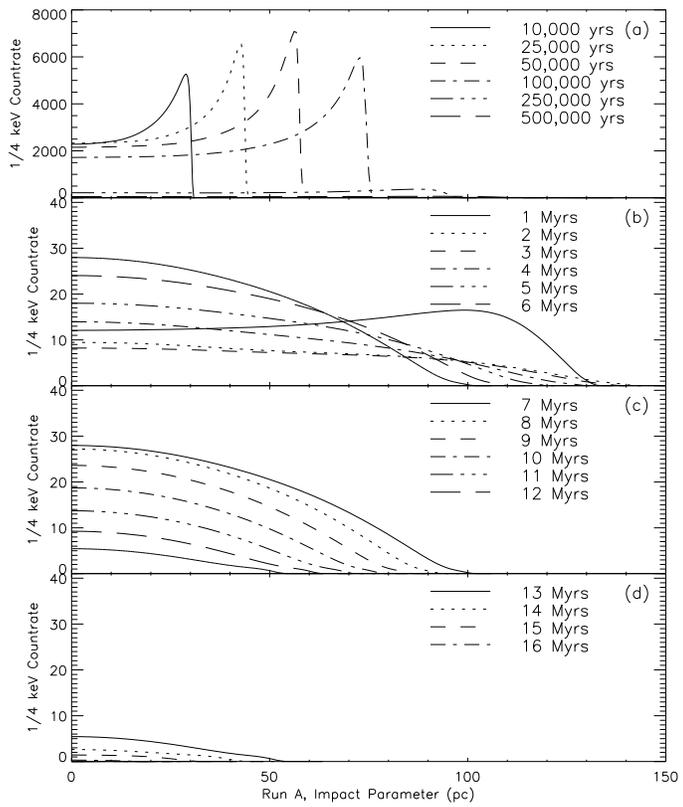
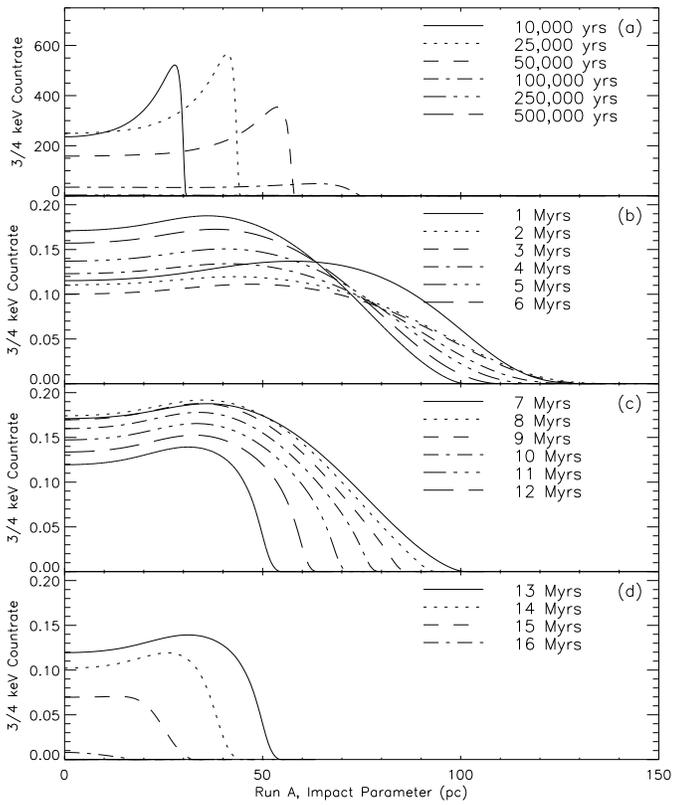
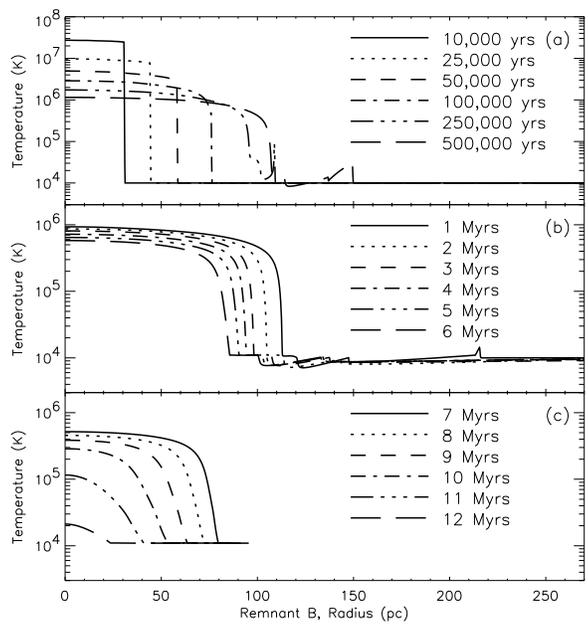
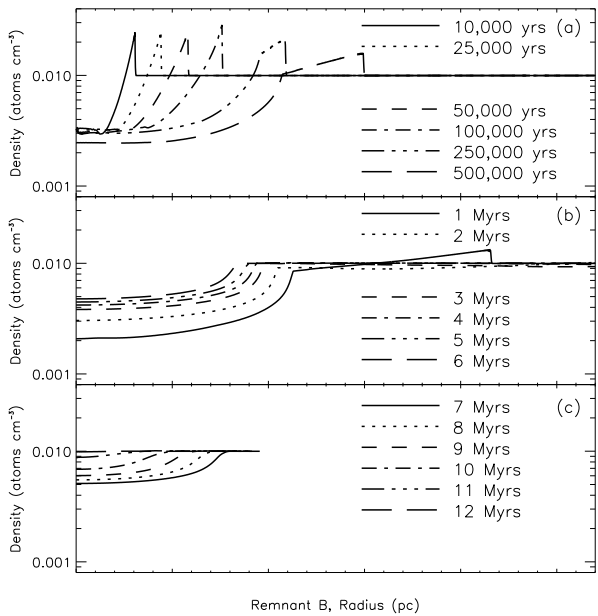
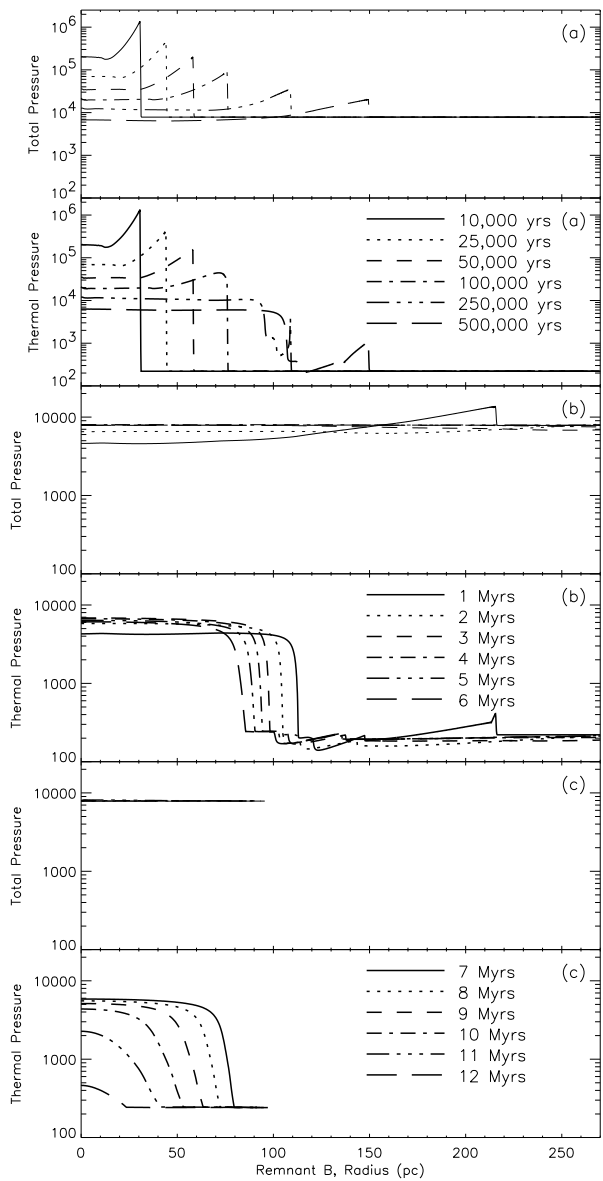
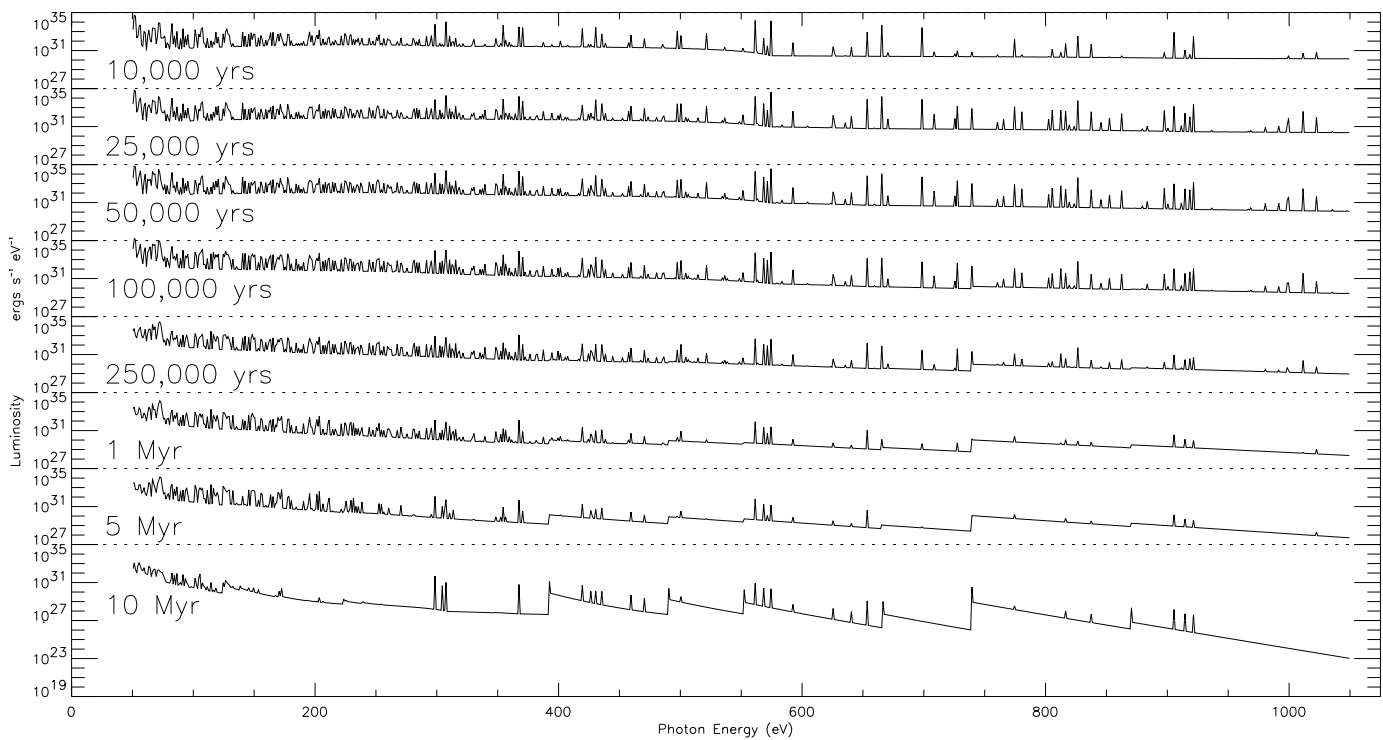
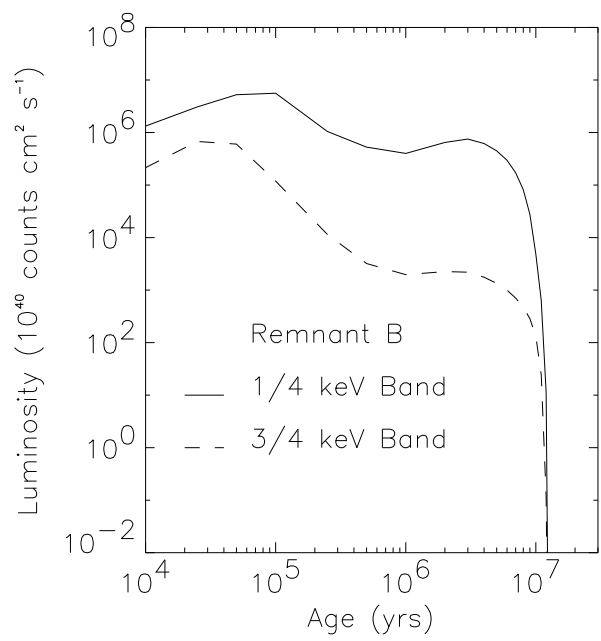
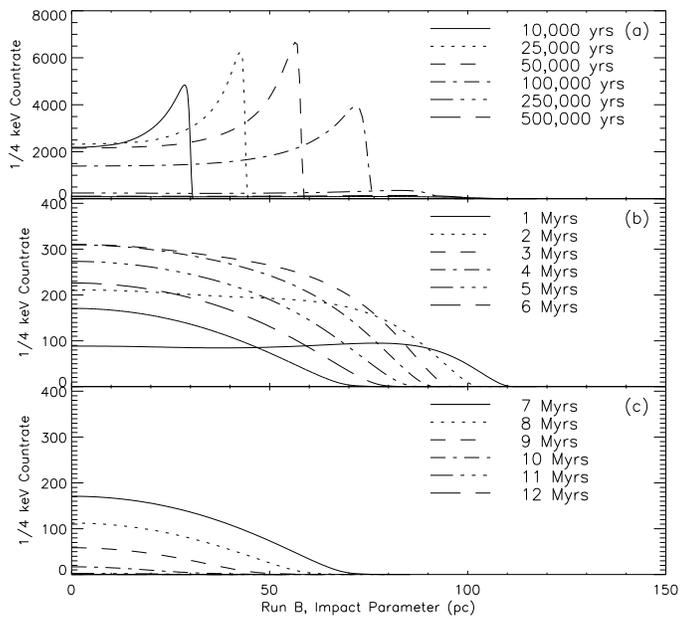
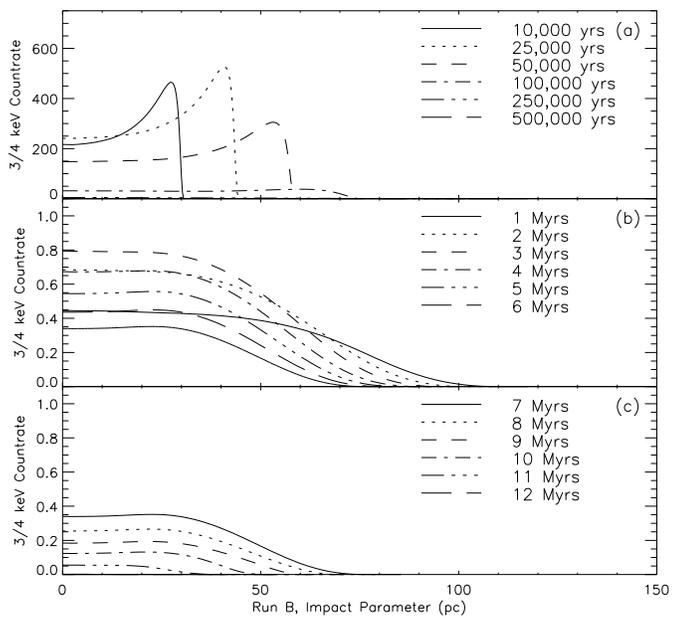
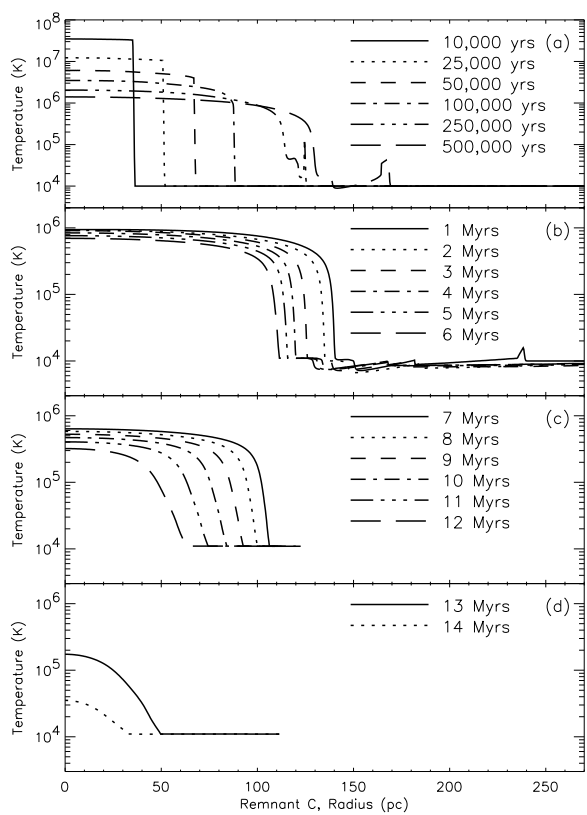
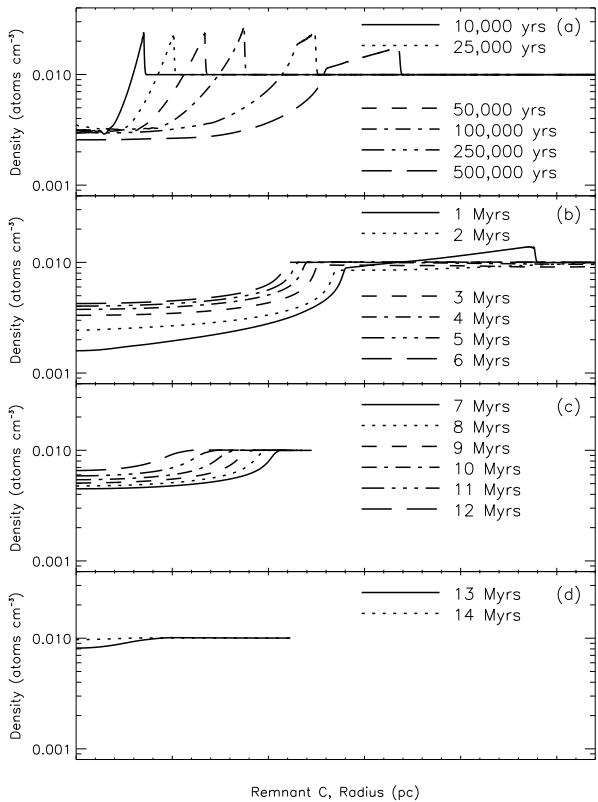
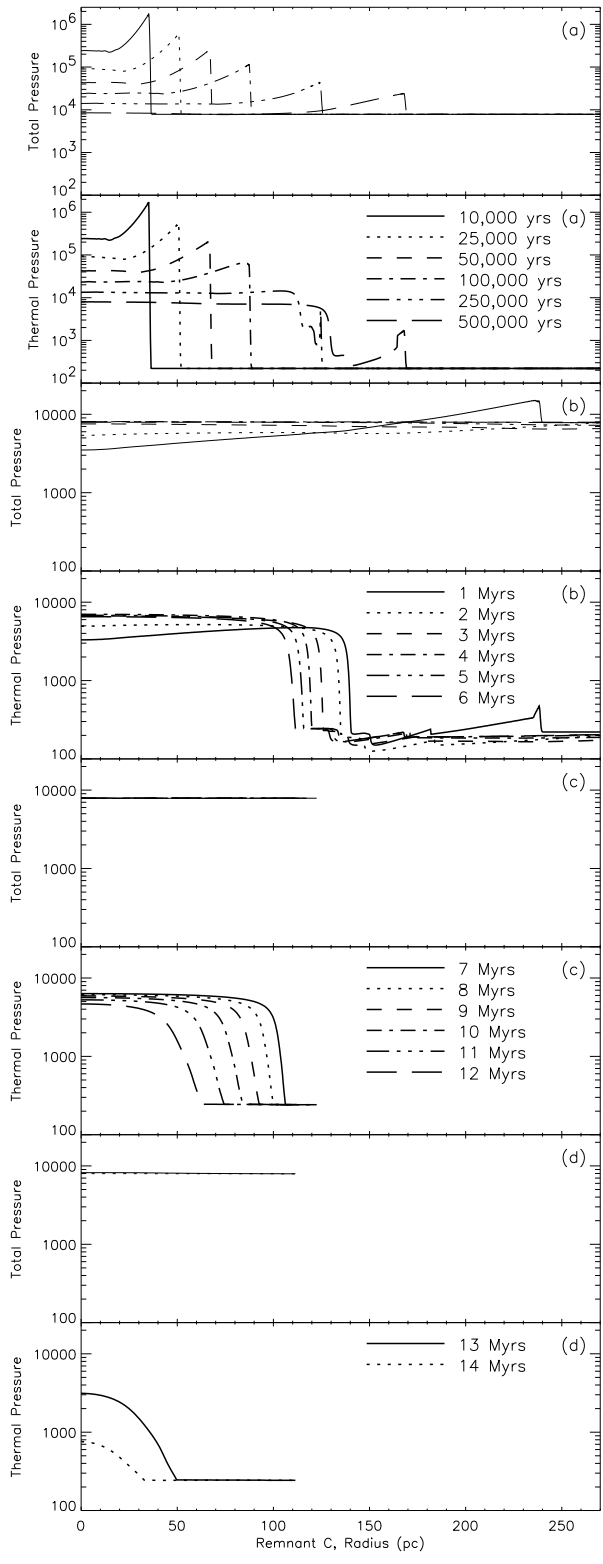
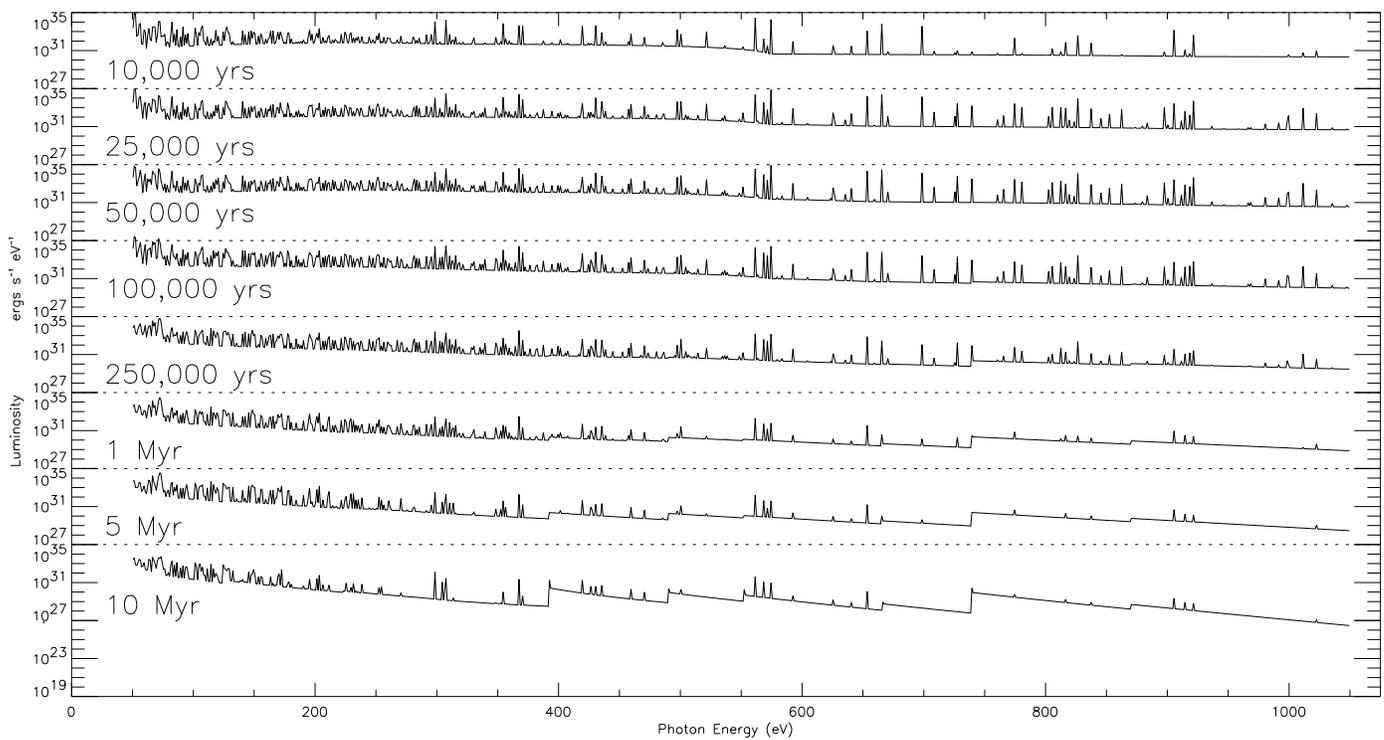
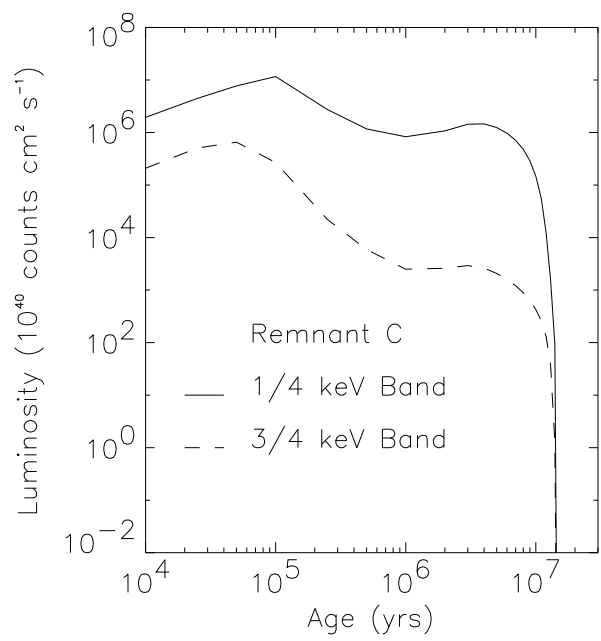
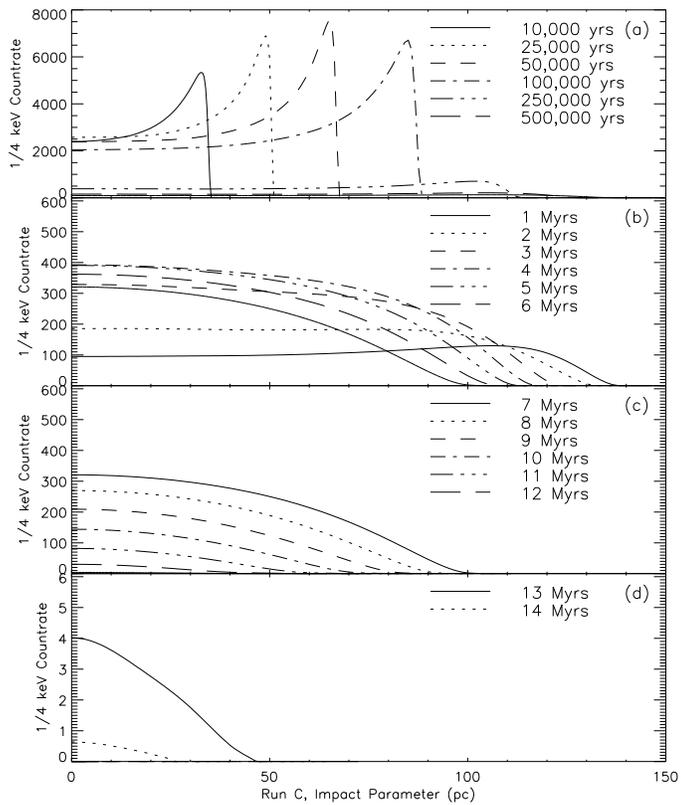
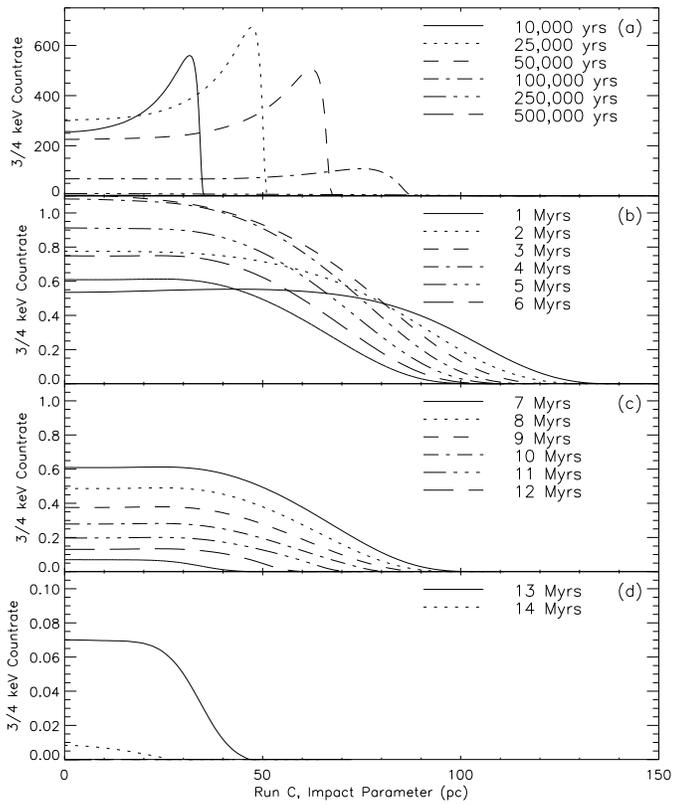
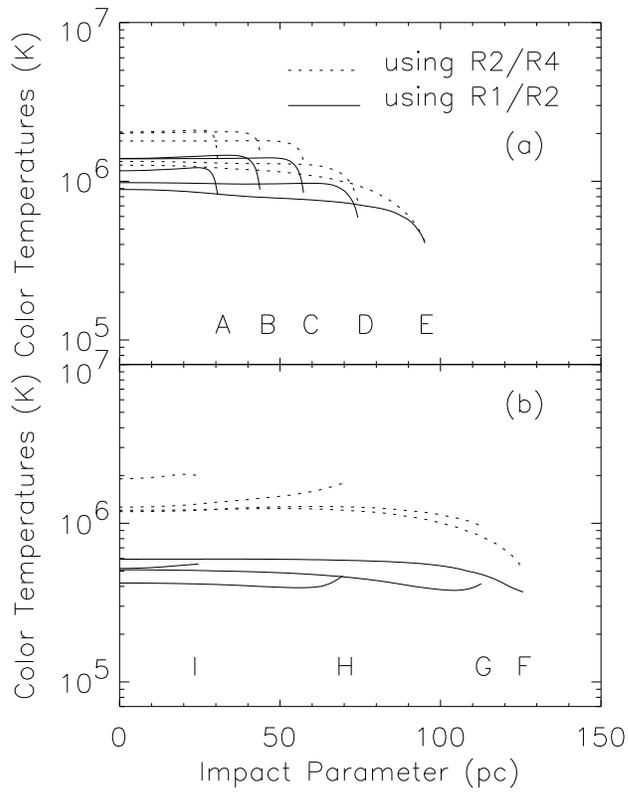
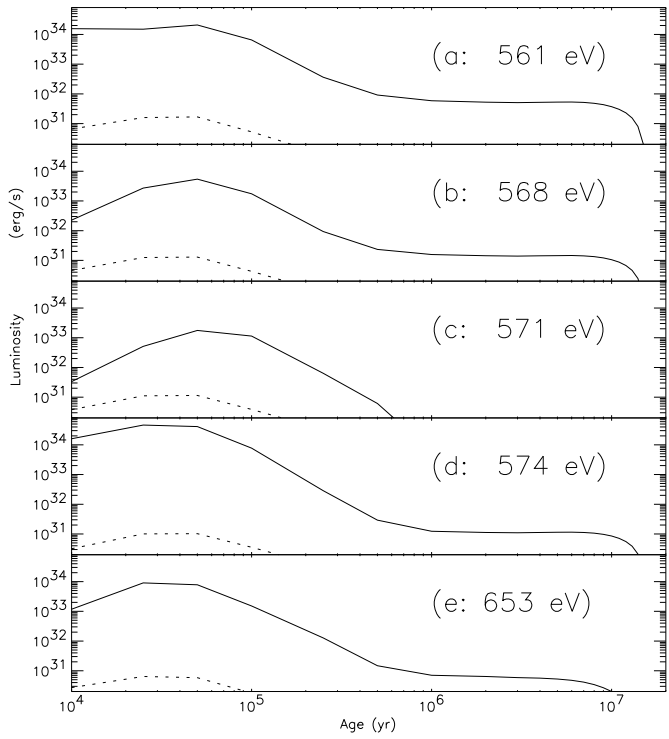
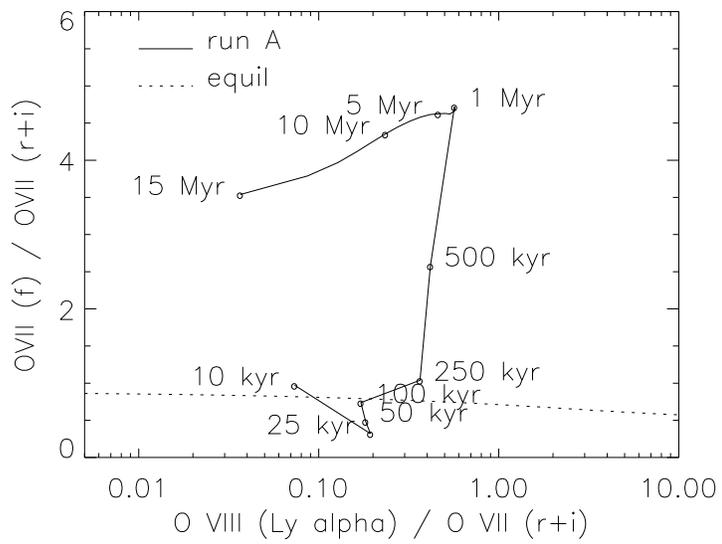
\end{document}